\documentclass[10 pt, showfigures]{natneu}
\pdfoutput=1
\usepackage[left=2.8cm,top=3cm,right=2.8cm]{geometry} 
\usepackage[pagewise]{lineno}

\usepackage{graphicx}
\usepackage{amsmath}
\usepackage{amssymb}
\usepackage{float}
\usepackage{color}
\usepackage{multicol}
\usepackage{xfrac}
\usepackage{xr}
\usepackage[unicode=true]{hyperref}
\newcommand{\overbar}[1]{\mkern 1.5mu\overline{\mkern-1.5mu#1\mkern-1.5mu}\mkern 1.5mu}
\usepackage{xcolor}
\hypersetup{
    colorlinks,
    linkcolor={red!50!black},
    citecolor={blue!50!black},
    urlcolor={blue!80!black}
}
\usepackage{bbold}

\newcommand*\patchAmsMathEnvironmentForLineno[1]{%
  \expandafter\let\csname old#1\expandafter\endcsname\csname #1\endcsname
  \expandafter\let\csname oldend#1\expandafter\endcsname\csname end#1\endcsname
  \renewenvironment{#1}%
     {\linenomath\csname old#1\endcsname}%
     {\csname oldend#1\endcsname\endlinenomath}}%
\newcommand*\patchBothAmsMathEnvironmentsForLineno[1]{%
  \patchAmsMathEnvironmentForLineno{#1}%
  \patchAmsMathEnvironmentForLineno{#1*}}%
\AtBeginDocument{%
\patchBothAmsMathEnvironmentsForLineno{equation}%
\patchBothAmsMathEnvironmentsForLineno{align}%
\patchBothAmsMathEnvironmentsForLineno{flalign}%
\patchBothAmsMathEnvironmentsForLineno{alignat}%
\patchBothAmsMathEnvironmentsForLineno{gather}%
\patchBothAmsMathEnvironmentsForLineno{multline}%
}

\title{Statistical mechanics of phase space partitioning \\ in large-scale spiking neuron circuits}
\author{Maximilian Puelma Touzel \textsuperscript{1,2,3,*} \&  Fred Wolf \textsuperscript{1,2,4} }
\begin{document}
\sloppy
\pagenumbering{gobble}

\maketitle
\begin{affiliations}
\item Max Planck Institute for Dynamics and Self-Organization, G{\"o}ttingen, Germany
\item Bernstein Center for Computational Neuroscience, G{\"o}ttingen, Germany
\item Laboratoire de Physique Th{\'e}orique, ENS-PSL Research University, Paris, France
\item Kavli Institute for Theoretical Physics, University of California Santa Barbara, Santa Barbara, USA
\end{affiliations}

\begin{frontpage}

\item[Corresponding Author:] ~\\
Maximilian Puelma Touzel\\
Laboratoire de Physique Théorique\\
Ecole Normale Supérieure\\
24 rue Lhomond\\
75231 Paris Cedex 05\\
Paris, France\\
Ph. +49 551 51 76 420\\
\textbf{e-mail:}~
\texttt{puelma@lpt.ens.fr}\\	

\end{frontpage}

\clearpage
\pagenumbering{arabic}

\begin{abstract}

Synaptic interactions structure the phase space of the dynamics of neural circuits and constrain neural computation. Understanding how requires methods that handle those discrete interactions, yet few exist.
Recently, it was discovered that even random networks exhibit dynamics that partitions the phase space into numerous attractor basins.
Here we utilize this phenomenon to develop theory for the geometry of phase space partitioning in spiking neural circuits. 
We find basin boundaries structuring the phase space are pre-images of spike-time collision events.
Formulating a statistical theory of spike-time collision events, we derive expressions for the rate of divergence of neighboring basins and for their size distribution. 
This theory reveals that the typical basin diameter grows with inhibitory coupling strength and shrinks with the rate of spike events. 
Our study provides an analytical and generalizable approach for dissecting how connectivity, coupling strength, single neuron dynamics and population activity shape the phase space geometry of spiking circuits.

\end{abstract}

\textbf{Key words:} \textit{neuronal circuits, dynamics of networks, disordered systems, basins of attraction, high dimensional systems, pulse-coupled systems, sequence generation}

\clearpage

\begin{introduction}

Computing devices, whether natural or artificial, perform their function by finely orchestrated state changes of internal dynamical variables.  In nervous systems these dynamical variables are physico-chemical states of nerve cells and synapses that connect them into complex networks called neural circuits.
The causal dependencies arising from the synaptic interactions between cells greatly extend the space of functions computable by the circuit, beyond that of single neurons. 

Mathematical models of neural circuits have been formulated in two fundamentally distinct ways\citep{Vogels2011}. 
Most synaptic interactions in the brain are driven by sparsely-fired nerve impulses, called spikes, each lasting only a millisecond. 
In spiking neural network models this fundamental granularity of neuronal interactions is explicitly represented: all interactions depend on a discrete set of spike event times.
Alternatively, continuous variable models for neural circuit dynamics are formulated by assuming that a frequency of nerve impulse generation, the firing rate, represents the information-encoding variable causally relevant for neural circuit computation. 
Firing rate models have been commonly used to model neural circuits \citep{Wilson1972}, theoretically study their dynamics \citep{Sompolinsky1988, Kadmon2015} and learning \citep{Hopfield1982, Sussillo2009, Laje2013,Brunel2016}, and are the basis of spectacular advances in artificial computing systems \citep{LeCun2015}. Statistical physics has played a role in this development, e.g. in clarifying the disordered phase space organization \citep{Gardner1999}. 

From a dynamical systems perspective, attractor states and their basins of attraction play a fundamental role in theories of neural computation. 
While analogous in some cases \citep{Harish2015,Mastrogiuseppe2017}, however, rate models are not equivalent to temporally coarse-grained versions of spiking neural networks, even if they are closely matched in structure \citep{Engelken2016}. 
Moreover, low firing rates (not much more than 1 Hz) in the cerebral cortex \citep{Roxin2011} make it hard to imagine how continuous rate variables associated to single neurons could provide a causally accurate description on behavioral time scales (hundreds of milliseconds). 
Developing theory for spiking networks may well require a dedicated approach. The absence of relevant averages and even a tractable ensemble of spiking trajectories, however, has thus far limited statistical approaches. Methods to design them \citep{ Memmesheimer2014} or to theoretically dissect the associated phase space organization are only starting to emerge. 

Recently it has been discovered that, with dominant inhibition, even randomly wired networks partition their phase space into a complex set of basins of attraction, termed \textit{flux tubes} \citep{Jahnke2008, Monteforte2012}.
Here we utilize this setting to develop a statistical theory of phase space partitioning in spiking neural circuits. We first present a simulation study of flux tubes, uncovering their shape and revealing it is structured by a spike time collision event. Formulating these events, we then derive the conditions for and rate of the mutual divergence of neighboring tubes. Our main calculation is the derivation of the distribution of flux tube sizes, which we obtain from statistics of these events by leveraging the random connectivity to average over the disorder. 
Our analytical approach provides a transparent method to determine how coupling strength, connectivity, single neuron dynamics and population activity combine to shape the phase space geometry of spiking neural circuits.

\end{introduction}

\section*{Methods}

We study a tractable instance of the inhibition-dominated regime of neural circuits. $N$ neurons are connected by an Erd\H{o}s-Rényi graph with adjacency matrix $A=\left(A_{mn}\right)$.
$A_{mn}=1$ denotes a connection from neuron $n$ to $m$, realized with probability, $p=K/N$. 
The neurons\rq{} membrane potentials, $V_{n}\in\left(-\infty,V_{T}\right]$, are governed by Leaky Integrate-and-Fire (LIF) dynamics,\vspace{5pt}
\begin{equation}
\tau\dot{V}_{n}(t)=-V_{n}(t)+I_{n}\left(t\right)\;,\label{eq:diffeq}
\end{equation}
for $n\in\left\{ 1,\dots,N\right\} $. Here, $\tau$ is the membrane time constant and $I_{n}\left(t\right)$ the synaptic current received by neuron $n$; when $V_{n}$ reaches a threshold, $V_{T}=0$, neuron $n$ `spikes\rq{}, and $V_{n}$ is reset to $V_{R}=-1$. At the spike
time, $t_{s}$, the spiking neuron, $n_{s}$, delivers a current pulse of strength $J$ to its $\mathcal{O}(K)$ postsynaptic neurons, $\left\{ m|A_{mn_{s}}=1\right\}$,  ($s$ indexes the spikes in the observation window). 
The total synaptic current is \vspace{5pt}
\begin{equation}I_{n}\left(t\right)=I_{\mathrm{Ext}}+\tau J\sum_{s}A_{nn_{s}}\delta(t-t_{s})\;,\label{eq:syncurrent}\end{equation}
where $I_{\mathrm{Ext}}>0$ is a constant external current and $J<0$ is the recurrent coupling strength. 
An $\mathcal{O}\left(1/\sqrt{K}\right)$-scaling of $J$ is chosen to maintain finite current fluctuations at large $K$ and implies that the external drive is balanced by the recurrent input. 
As a consequence, firing in this network is robustly asynchronous and irregular \citep{VanVreeswijk1996,Brunel1999,Renart2010,Barral2016}. 
Setting $I_{\mathrm{Ext}}=\sqrt{K}I_{0}$, with $I_{0}>0$, and $J=-J_{0}/\sqrt{K}$ with $J_{0}>0$, the corresponding stationary mean-field equation for the population-averaged firing rate, $\bar{\nu}$, is \citep{Monteforte2012}\vspace{5pt}
\begin{equation}
\bar{\nu}=\frac{I_{0}}{J_{0}\tau}+\mathcal{O}\left(\frac{1}{\sqrt{K}}\right)\;.\label{eq:bal eqn-11}
\end{equation}
It is convenient to map the voltage dynamics to a pseudophase representation \citep{Jin2002,Monteforte2012}, $\vec{\phi}\left(t\right)$, with\vspace{5pt}
\begin{equation}
\phi_{n}(t)=\frac{\tau}{T_{\mathrm{free}}}\ln\left[\frac{I_{\mathrm{Ext}}-V_{R}}{I_{\mathrm{Ext}}-V_{n}(t)}\right]\;,\label{eq:volt to phase}
\end{equation}
where $T_{free}$ is the oscillation period of a neuron driven only by $I_{\mathrm{Ext}}$. $\phi_{n}\left(t\right)$ evolves linearly in time,\vspace{5pt} 
\begin{equation}
\dot{\phi}_{n}\left(t\right)=T_{free}^{-1}\;,\label{eq:psuedophase}
\end{equation}
between spike events, \emph{i.e.} $t\notin\left\{ t_{s}\right\} $, and undergoes shifts given by the phase response curve, $Z(\phi)$, across input spike times where $\phi$ is the state at spike reception. 
In the large-$K$ limit, $T_{\mathrm{free}}$ and $Z\left(\phi\right)$ simplify to \vspace{5pt}
\begin{eqnarray}
T_{\mathrm{free}} & \approx & \frac{\tau}{I_{\mathrm{Ext}}}=(\sqrt{K}J_{0}\bar{\nu})^{-1}\;,\label{eq:large K Tfree =000026 d}\\
Z\left(\phi\right) & \approx & -d\phi+const.\nonumber \\
\textrm{with}\;\;d: & = & \frac{\left|J\right|}{I_{\mathrm{Ext}}}=(K\bar{\nu}\tau)^{-1}\;,\label{eq:dapprox}
\end{eqnarray}
respectively (see Supplemental Methods for details). 
The differential phase response, $\tfrac{\mathrm{d}}{\mathrm{d}\phi}Z=-d$, is essential for the strongly dissipative nature of the collective dynamics. For
$J=0$, the dynamics (equation \eqref{eq:psuedophase}) would preserve phase space volume. This volume, however, is strongly contracted by spikes received in the post-synaptic neurons. Consider trajectories from a small ball of initial conditions as they emit the same future spike. The ball of phases at this spike contracts by a factor $1-d$ along each of the $K$ dimensions of the subspace spanned by the post-synaptic neurons. The volume thus contracts by $(1-d)^K\to e^{\lambda_{\mathrm{inh}}}$ per spike, for $K\gg 1$, with exponential rate, \vspace{5pt}
\begin{equation}
\lambda_{\mathrm{inh}}\approx-Kd\;.\label{eq:conrate}
\end{equation}
$\lambda_{\mathrm{inh}}<0$ is responsible for the linear stability of the dynamics given by equations \eqref{eq:diffeq} and \eqref{eq:syncurrent}, first shown in Refs. \citep{Jin2002,Jahnke2008}.

\section*{Phase-space partitioning}

\begin{figure}[h]
 \centering
   \includegraphics[width=0.9\textwidth]{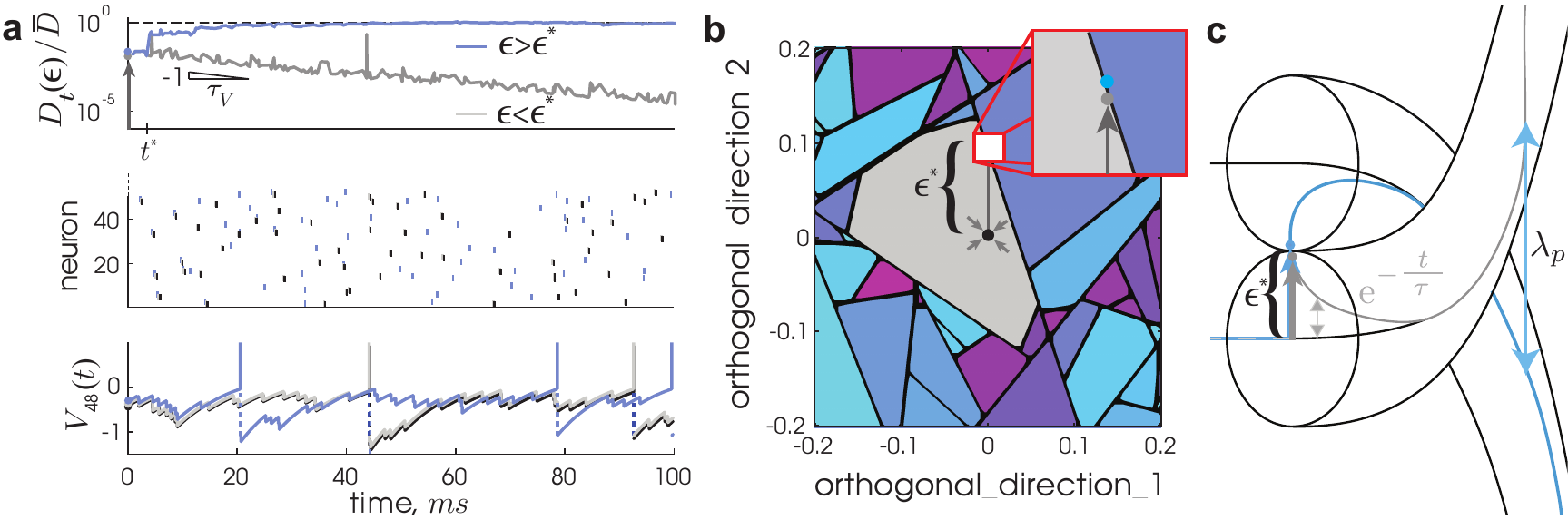}
\caption{\small{
Finite-size perturbation instability and phase space partitioning in spiking networks. 
The three panels display the same slightly subcritical and supercritical perturbation of strength ${\epsilon}^{*} \pm \delta$, $\delta \gtrsim 0$, respectively, applied once at $t=0$ and in a random direction away from an equilibriated trajectory. 
\textbf{(a)} Temporal responses of the system. 
\emph{Top}: The corresponding distance time series, $D_{t}(\epsilon)$, between the perturbed and unperturbed
trajectories (gray: sub-critical, blue: super-critical; arrows in all three panels indicate the respective the perturbation). 
The divergence of $D_{t}(\epsilon^{*}+\delta)$ begins at $t^{*}\approx 3\;ms$, and saturates at the average distance between randomly chosen trajectories, $\bar{D}$ (dashed line) \citep{Monteforte2012}, while $D_{t}(\epsilon^{*}-\delta)$ only decays exponentially. 
\emph{Middle}: The spike times as vertical ticks of the first 50 randomly labeled neurons from the network. The unperturbed sequence is shown in black. 
\emph{Bottom}: The subthreshold voltage time course of an example neuron. 
The spike sequence and membrane potentials of the sub and supercritical trajectories decorrelate after $t^{*}$. 
\textbf{(b)} A 2D cross-section $\left(\delta\phi_{1},\delta\phi_{2}\right)$ of the pseudophase representation of the phase space, orthogonal to and centered on the unperturbed trajectory from (a) at $t=0$ (see also Ref. 15). 
The black dot at the origin indicates the latter, whose attractor basin is colored gray. 
The other colors distinguish basins in the local neighborhood. 
The two perturbed trajectories from (a) were initiated from $\left(\delta\phi_{1},\delta\phi_{2}\right)=\left(0,\epsilon^{*}\pm\delta\right)$, respectively (shown as gray and blue dots, respectively, in the inset, in (a,Top and Bottom), and in (c)). 
\textbf{(c)} Schematic phase space caricature of two neighboring flux tubes with subcritical perturbations decaying on the order of the membrane time constant, $\tau$, and typical basin diameter, $\epsilon^{*}$. 
The pseudoLyapunov exponent, $\lambda_{p}$, is the rate at which neighboring tubes separate from each other (parameters: $N=200$, $K=50$, $\bar{\nu}=10\mbox{\ Hz}$, $\tau=10\mbox{\ ms}$, $J_{0}=1$).}}
\label{fig:The-known-LIF}
\end{figure}

The phase space volume taken up by an ensemble of nearby trajectories at a given spike is contracted at the spike\rq{}s reception. 
Larger phase space volumes, however, are not uniformly contracted but torn apart, with the pieces individually contracted and overall dispersed across the entire traversed phase space volume. 
The elementary phenomenon is illustrated in Fig. \ref{fig:The-known-LIF}. 

We define the \emph{critical} \emph{perturbation strength}, $\epsilon^{*}$, as the flux tube\rq{}s extent out from a given state $\vec{\phi}_{0}$ on the equilibriated trajectory, $\vec{\phi}_{t}$, and in a given orthogonal perturbation direction, $\vec{\xi}$, \vspace{5pt}
\begin{equation}
\epsilon^{*}(\vec{\phi}_{0},\vec{\xi}):=\sup\left\{ \epsilon\left|\lim_{t\to\infty}D_{t}\left(\epsilon\right)=0\right.\right\} \;.\label{eq:epscritdef}
\end{equation}
Here, $D_{t}\left(\epsilon\right)$ is the 1-norm distance, \vspace{5pt}
\begin{equation}
D_{t}\left(\epsilon\right):=\frac{1}{N}\sum_{n=1}^{N}\left|\phi_{n,t}-\phi_{n,t}\left(\epsilon\right)\right|,
\end{equation}
between $\vec{\phi}_{t}$ and the perturbed trajectory, $\vec{\phi}_{t}(\epsilon)$, evolving freely from the perturbed state, $\vec{\phi}_{0}\left(\epsilon\right):=\vec{\phi}_{0}+\epsilon\vec{\xi}$ (reference time $t=0$ and $||\vec{\xi}||=1$; see Supplemental Methods for details). 
 $\epsilon^{*}$ is the largest value below which $D_{t}\left(\epsilon\right)$ vanishes in time. 
$D_{t}$ initially decays exponentially, but for a supercritical perturbation, $\epsilon>\epsilon^{*}$, there exists a \emph{divergence event time,} $t^{*}>0$, defined and obtained as the time at which a sustained divergence in $D_{t}$ begins (see Fig. \ref{fig:The-known-LIF}a).

A 2D cross-section of the phase space around $\vec{\phi}_{0}$ (Fig. \ref{fig:The-known-LIF}b) reveals that the locations of these critical perturbations form lines which intersect to form polygon-shaped basin boundaries. 
Before developing a theory for this phase space organization (caricatured in Fig. \ref{fig:The-known-LIF}c), we first analyze two main features of the geometry of a flux tube: the punctuated exponential decay of its cross-sectional volume and the exponential separation of neighboring tubes.

\section*{Punctuated geometry of flux tubes}

\begin{figure}[t]
\includegraphics{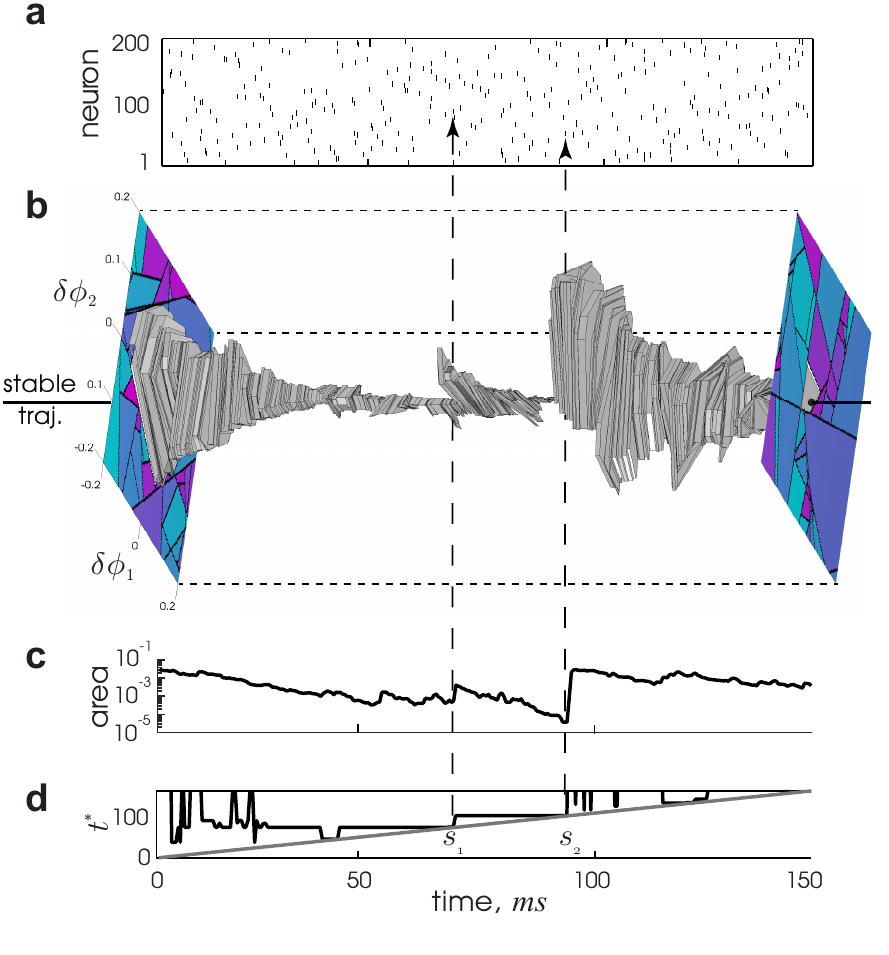}
\centering{}
\caption{
The basin boundary contracts towards and can blowup away from the stable trajectory within it. 
\textbf{(a)} Spike times from all neurons of the simulated trajectory, $\vec{\phi}_{t}$, in a $150$ ms window.
\textbf{(b)} 2+1D folded phase space volume, $\left(\delta\phi_{1},\delta\phi_{2},t\right)$, centered around $\vec{\phi}_{t}$ located at $\left(0,0,t\right)$ (black line) and extended in two fixed, random directions, $\vec{\delta\phi_{1}}$ and $\vec{\delta\phi_{2}}$. 
The center tube is filled gray in this volume, and the two cross-sections, $\left(\delta\phi_{1},\delta\phi_{2},0\right)$ and $\left(\delta\phi_{1},\delta\phi_{2},150\right)$, are shown.
\textbf{(c)} Cross-sectional area of the center tube from (b) versus time. 
The area decays exponentially but can undergo abrupt expansions at blow-up times, e.g. at spikes $s_{1}$ and $s_{2}$ (note the logarithmic scale on the ordinate). 
\textbf{(d)} The absolute time of the next divergence event, $t^{*}$ (see Fig. \ref{fig:The-known-LIF}a, top), versus time, for perturbations along $\vec{\delta\phi_{1}}$. 
Note the step increase coincident with the blowup events seen in (b,c) (vertical, dashed lines). (Same parameters as Fig. \ref{fig:The-known-LIF}.)}
\label{fig:Tube-evolution-is}
\end{figure}

As expected from the typical phase space volume contraction (see equation \eqref{eq:conrate}), we find along a simulated trajectory that the orthogonal phase space volume enclosed by the local flux tube exhibits exponential decay. 
This decay, however, is punctuated by blowup events. Figure \ref{fig:Tube-evolution-is}a displays the spiking activity produced by the typical trajectory, $\vec{\phi}_{t}$. 
The neighborhood around $\vec{\phi}_{t}$ over a time window is visualized in a \emph{folded representation} using a fixed, 2D projection of the phase space (Fig. \ref{fig:Tube-evolution-is}b and \href{file:http://www.phys.ens.fr/~mptouzel/pdf/PuelmaTouzel_Partioning_suppvideo.avi}{Supplementary Video}; see Supplemental Information for construction details).
The basin of attraction surrounding $\vec{\phi}_{t}$ (Fig. \ref{fig:Tube-evolution-is}b) consists of lines which remain fixed between spike times. 
Across spike times, new lines appear and existing lines disappear. 
At irregular intervals breaking up time windows of exponential contraction, large abrupt blowup events take the boundary away from the center trajectory (Fig. \ref{fig:Tube-evolution-is}b, c), producing jumps in the area enclosed by the boundary. 
It is important to note that these events do not mean that the evolving phase space volume from an ensemble of states contained in the tube would expand. 
Such volumes only contract and converge to the same asymptotic trajectory. 
The basin of attraction itself, however, does not exclusively contract with time. 
In fact, it should on average maintain a typical size.

The blowup events typically coincide with a divergence event time, $t^{*}$ (Fig. \ref{fig:The-known-LIF}a), in some perturbation direction.
Two such coincidences are visible in Fig. \ref{fig:Tube-evolution-is}c,d. We conclude that the local basin at any time extends out in phase space until the perturbed trajectory approaches the pre-image of a divergence event occurring at a future time. 
Flux tube shape is then determined by the statistics of such events.

\section*{Tube boundary and divergence}

We analyzed a set of divergence events from simulations. We find that a collision of a pair of spikes constitutes the elementary event triggering the divergence of the perturbed trajectory. These pairs, hereon called \emph{susceptible} spike pairs, were generated by connected pairs of neurons. 
Moreover, a perturbation-induced collision of a susceptible spike pair generated an abrupt spike time shift in one or both of the connected neurons\rq{} spike times. 
We found that the nature of the spike time shift depends on the motif by which the two neurons connect. 
We denote the \emph{backward-connected} pair motif $n_{s^{*}}\leftarrow n_{s\rq{}}$, where $s^{*}$, the \emph{decorrelation event index}, is the spike index of the earlier of the pair (note that $t^{*}\equiv t_{s^{*}}$), and $s\rq{}>s^{*}$ labels the later spike in the pair.

\begin{figure}[t]
\includegraphics{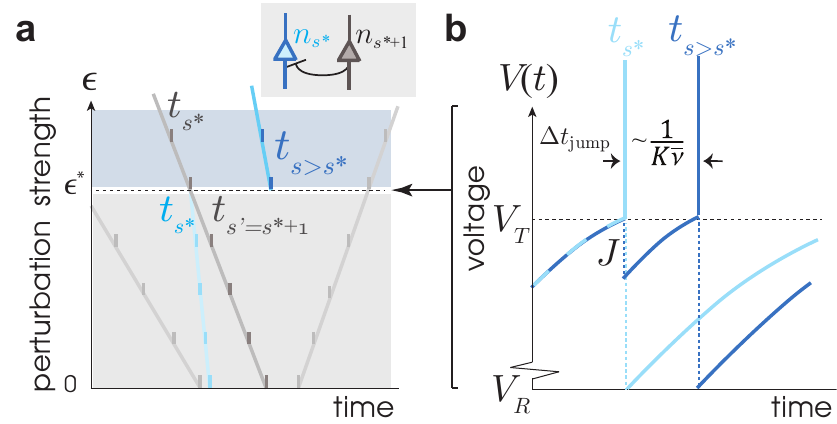}
\centering{}
\caption{
The collision of a susceptible spike pair causes an abrupt change in spike time.
\textbf{(a)} A schematic illustration of the collision event for the backward-connected pair motif (shown in inset). 
For this motif, the interval vanishes as $\epsilon\to\epsilon^{*}$ from below.
Perturbation strength, $\epsilon$, is plotted versus time, where the timings of spikes for every perturbation strength are indicated as ticks on lines. 
The spike times shift continuously for $\epsilon<\epsilon^{*}$. 
As the next input spike time, $t_{s^{*}+1}\left(\epsilon^{*}-\delta\right)$, is advanced over $t_{s^{*}}\left(\epsilon^{*}+\delta\right)$A discontinuous jump of size $\Delta t_{\mathrm{jump}}$ occurs in the spike time of the post-synaptic neuron, $n_{s^{*}}$ (light to dark blue) from $t_{s^{*}}\left(\epsilon^{*}-\delta\right)$ to $t_{s^{\rq{}}>s^{*}}\left(\epsilon^{*}+\delta\right)$, $\delta \gtrsim 0$. 
\textbf{(b)} Schematic illustration of the voltage of the $n_{s^{*}}$ neuron versus time for $\epsilon^{*}\pm\delta$. 
The inhibitory kick of size $J=-J_{0}/\sqrt{K}$ (not shown to scale) delays the spike time by an amount $\Delta t_{\mathrm{jump}}\sim\left(K\bar{\nu}\right)^{-1}$.
}
\label{fig:Discontinuous-jumps-in-1}
\end{figure}

For $\epsilon \lesssim \epsilon^{*}$, the presynaptic spike time, $t_{s\rq{}=s^{*}+1}$, is advanced with increasing $\epsilon$ relative to the postsynaptic spike time, $t_{s^{*}}$, until the two spikes collide (see Fig. \ref{fig:Discontinuous-jumps-in-1}).
At collision ($\epsilon=\epsilon^{*}$), the pulsed inhibition and the rate of approach to voltage threshold cause an abrupt delay of $t_{s^{*}}$ by $\Delta t_{\mathrm{jump}}$. 
Using equation \eqref{eq:bal eqn-11}, we obtain \vspace{5pt}
\begin{align}
\Delta t_{\mathrm{jump}} & =\tau \ln \left[1+d\right]\approx\tau d=\left(K\bar{\nu}\right)^{-1}\;\label{eq:Tjump}
\end{align}
for $d\ll 1$. 
Further details and the other two motifs (forward-connected and symmetrical) are discussed in the Supplementary Notes. 

For each spike in the network sequence, the rate of its susceptible spike partners is \vspace{5pt}
\begin{equation}
\lambda_{\mathrm{sus}}=p/\overbar{\Delta t}=K\bar{\nu}\;,\label{eq:tsus}
\end{equation}
where $\overbar{\Delta t}=\left(N\bar{\nu}\right)^{-1}$ is the average distance between successive spikes. 
Since $\Delta t_{\mathrm{jump}}\approx \lambda_{\mathrm{sus}}^{-1}$, the spike time of neuron $n_{s^*}$ is shifted forward typically as far as its next nearest susceptible partner spike. 
Thus, one collision event will typically induce another in at least one of the $\mathcal{O}\left(K\right)$ neurons to which the involved pair of neurons are presynaptic. 
A cascade of collision events then follows with near certainty (see Supplemental Notes for details). 

The shift in $t_{s^{*}}$ by $\Delta t_{\mathrm{jump}}$ is carried forward to all future spike times of $n_{s^{*}}$, so that $n_{s^{*}}$ becomes a source of collision events. 
The total collision rate is then $\lambda_{\mathrm{sus}}$ multiplied by the number of source neurons, which approximately increments with each collision in the cascade. 
Averaging over realizations of the cascade (reference time $t^{*}=0$), the average number of collisions, $\bar{m}$, grows as $\tfrac{\mathrm{d}}{\mathrm{d}t}\bar{m}=\lambda_{\mathrm{sus}}\bar{m}$. 
Finally, since each collision produces a jump in distance of equal size, we obtain the pseudoLyapunov exponent, $\lambda_{p}=\lambda_{\mathrm{sus}}$ from its implicit definition, $\bar{D}_{t}= \bar{D}_{0}\exp\left[\lambda_{p}t\right]$ (see Supplemental Notes), as the exponential rate at which flux tubes diverge.

\section*{Statistical theory of flux tube diameter}

The geometry of a flux tube is captured by the \emph{flux tube indicator function}, $\mathbb{1}_{\mathrm{FT}}\left(\epsilon\right)=\Theta\left(\epsilon^{*}(\vec{\phi}_{0},\vec{\xi})-\epsilon\right)$, evaluated across network states, $\vec{\phi}_{0}$, of its contained attracting trajectory and perturbation directions, $\vec{\xi}$. 
Using the Heaviside function, $\Theta(x)$, $\mathbb{1}_{\mathrm{FT}}\left(\epsilon\right)=1$ for perturbations remaining in the tube ($\epsilon<\epsilon^{*}$), and 0 otherwise.
The average of $\mathbb{1}_{\mathrm{FT}}\left(\epsilon\right)$ over $\vec{\phi}_{0}$ and $\vec{\xi}$ , \vspace{5pt}
\begin{equation}
\hat{S}\left(\epsilon\right)=\left[\mathbb{1}_{\mathrm{FT}}\left(\epsilon\right)\right]_{\rho\left(\vec{\phi}_{0},\vec{\xi}\right)}\;,\label{eq:fracrestore}
\end{equation}
is the \emph{survival} \emph{function}: the probability that an $\epsilon$-sized perturbation does not lead to a divergence event later in the perturbed trajectory, \textit{i.e.} $\epsilon<\epsilon^{*}$, and is formally defined as $\hat{S}\left(\epsilon\right):=1-\int_{0}^{\epsilon}\rho\left(\epsilon^{*}\right)\mbox{d}\epsilon^{*}$, with $\rho\left(\epsilon^{*}\right)$ the transformed density over $\epsilon^*$. 
$\hat{S}\left(0\right)=1$ and decays to 0 as $\epsilon\to\infty$. 
The scale of this decay defines the typical flux tube size. 
Calculating $\hat{S}\left(\epsilon\right)$ requires two steps: firstly, establishing a tractable representation of $\epsilon^{*}(\vec{\phi}_{0},\vec{\xi})$ and secondly, performing the average in equation \eqref{eq:fracrestore}. 
Both of these in general pose intricate problems. However, as we will see next, both substantially simplify when generic properties of the asynchronous, irregular state are taken into account. 

Perturbed spike intervals are obtained using the spike time deviations, $\delta t_{s}\left(\epsilon\right):=t_{s}\left(\epsilon\right)-t_{s}\left(0\right)$, $s=1,2,\dots$,\vspace{5pt}
\begin{align}
\Delta t_{s}\left(\epsilon\right) = t_{s}(\epsilon)-t_{s-1}(\epsilon)=\Delta t_{s}\left(0\right)+\delta t_{s}\left(\epsilon\right)-\delta t_{s-1}\left(\epsilon\right)\,,\;s\geq 2.\label{eq:deltat}
\end{align}
In a linear approximation we find,\vspace{5pt}
\begin{equation}
\delta t_{s}\left(\epsilon\right)\approx-Ca_{s}\epsilon\;,\label{eq:smallepsapprox}
\end{equation}
where $C=\frac{T_{\mathrm{free}}}{\sqrt{N}}$ converts network phase deviation to spike time deviation and $a_{s}$ is a dimensionless susceptibility that depends on the adjacency matrix, $A=\left(A_{mn}\right)$, derivatives of the phase response curve evaluated at the network states at past spike times, $\{\vec{\phi}_{s'}=\vec{\phi}_{t_{s'}}\}$ for $s'<s$, and the perturbation direction $\vec{\xi}$ (see Supplemental Notes
for its derivation). 
Substituting equation \eqref{eq:smallepsapprox} into equation \eqref{eq:deltat} gives\vspace{5pt} 
\begin{equation}
\Delta t_{s}\left(\epsilon\right)\approx\Delta t_{s}-C\Delta a_{s}\epsilon\;,\label{eq:pertint}
\end{equation}
with $\Delta t_{s}=\Delta t_{s}\left(0\right)$. 
Note that $\Delta t_{s}\left(\epsilon\right)$ can have a zero, \emph{i.e.} a spike time collision only when $\Delta a_{s}=a_{s}-a_{s-1}>0$.

To obtain the scaling behavior of the flux tube geometry it is sufficient to examine the statistics of flux tube borders using the corresponding divergence events generated by collisions of backward-connected susceptible spike pairs in the perturbed trajectory (Fig. \ref{fig:Discontinuous-jumps-in-1}). 
In these cases, the perturbation strength $\epsilon\to\epsilon^{*}$ as the network spike interval $\Delta t_{s^*}\left(\epsilon\right)\to0$ for $A_{n_{s^*}n_{s^*+1}}=1$. 
In fact, the latter condition serves in these cases as an implicit definition of $\epsilon^*$ and $s^*$.

According to equation \eqref{eq:pertint}, $\hat{S}\left(\epsilon\right)$ in principle depends on the adjacency matrix, $A=\left(A_{mn}\right)$, of the network realization. 
Removing this dependence by averaging over the ensemble of graphs, $P_{A}\left(\left(A_{mn}\right)\right)$, simplifies the calculation of the survival function,\vspace{5pt}
\begin{equation}
S\left(\epsilon\right)=\left[\hat{S}\left(\epsilon\right)\right]_{P_{A}\left(\left(A_{mn}\right)\right)}\;.\label{eq:avggraph}
\end{equation}
Evaluating the right-hand side of equation \eqref{eq:avggraph} using the perturbed spike intervals, linearized in $\epsilon$, requires knowledge of the joint probability density of all variables present in equation \eqref{eq:pertint}, \vspace{5pt}
\begin{equation}
\rho_{T}=\rho(\left\{ \Delta a_{s}\right\} ,\left\{ \Delta t_{s}\right\} ,M,\vec{\phi}_{0}\vert \; \vec{\xi}, \left(A_{mn}\right))\;\rho(\vec{\xi})\;P_{A}\left(\left(A_{mn}\right)\right),\label{eq:prob}
\end{equation}
where we have chosen the perturbation direction, $\vec{\xi}$, to be statistically independent of the state, $\vec{\phi}_{0}$, being perturbed at $t=0$.
Here, the unperturbed spike pattern is represented by two random variables: $M$, the number of spikes in the time interval $\left[0,T\right]$ after the perturbation, and $\left\{ \Delta t_{s}\right\} $, the set of all $M-1$ inter-spike intervals in this window. 
It is well understood that in the large-system limit in a sparse graph, $1\ll K\ll N$, the currents driving individual neurons in the network converge to independent, stationary Gaussian random functions \citep{Tuckwell1988}. 
For low average firing rates, this implies that the pattern of network spikes $\left(M,\left\{ \Delta t_{s}\right\} \right)$ resembles a Poisson process \citep{Lindner2006}. 
Furthermore, the susceptibility becomes state-independent in this limit. 
Neglecting the weak dependence between the distribution of network spike patterns and $A=\left(A_{mn}\right)$, the full density, $\rho_{T}$ (equation \eqref{eq:prob}), approximately factorizes, \vspace{5pt}
\begin{equation}
\rho_{T}\sim P_{A_{mn}}\left(A_{mn}\right)P_T(M)\prod_{s=2}^{M}\rho_{t}\left(\Delta t\right)2\Theta(\Delta a_{s}) \rho_{a}\left(\Delta a_{s}\right)\;,\label{eq:decomrho}
\end{equation}
with distribution of a single adjacency matrix element, $P_{A_{mn}}\left(A_{mn}=1\right)=p$, $P_{A_{mn}}(A_{mn}=0)=1-p$, count distribution of spikes in the observation window, $P_T(M)$, and distribution of single inter-spike interval $\rho_{t}\left(\Delta t\right)$. The latter is exponential with rate $\overbar{\Delta t}$.
With these approximations (see Supplementary Notes for details), all dependencies on the distribution of perturbation direction are mediated by the susceptibilities, $\left\{ \Delta a_{s}\right\} $.
For any isotropic $\rho(\vec{\xi})$ having finite-variance, $\rho_{a}\left(\Delta a_{s}\right)$ has zero mean and standard deviation proportional to $\exp \left[ \frac{\lambda_{\mathrm{inh}}}{N}s\right]$, with the average contraction rate per neuron, $\frac{\lambda_{\mathrm{inh}}}{N}=-\frac{Kd}{N}=-pd$, due to the inhibition. The factor $2\Theta(\Delta a_{s})$ places support only  positive values of $\Delta a_{s}$ as required.

\begin{figure}[ht!]
\noindent \begin{centering}
\includegraphics{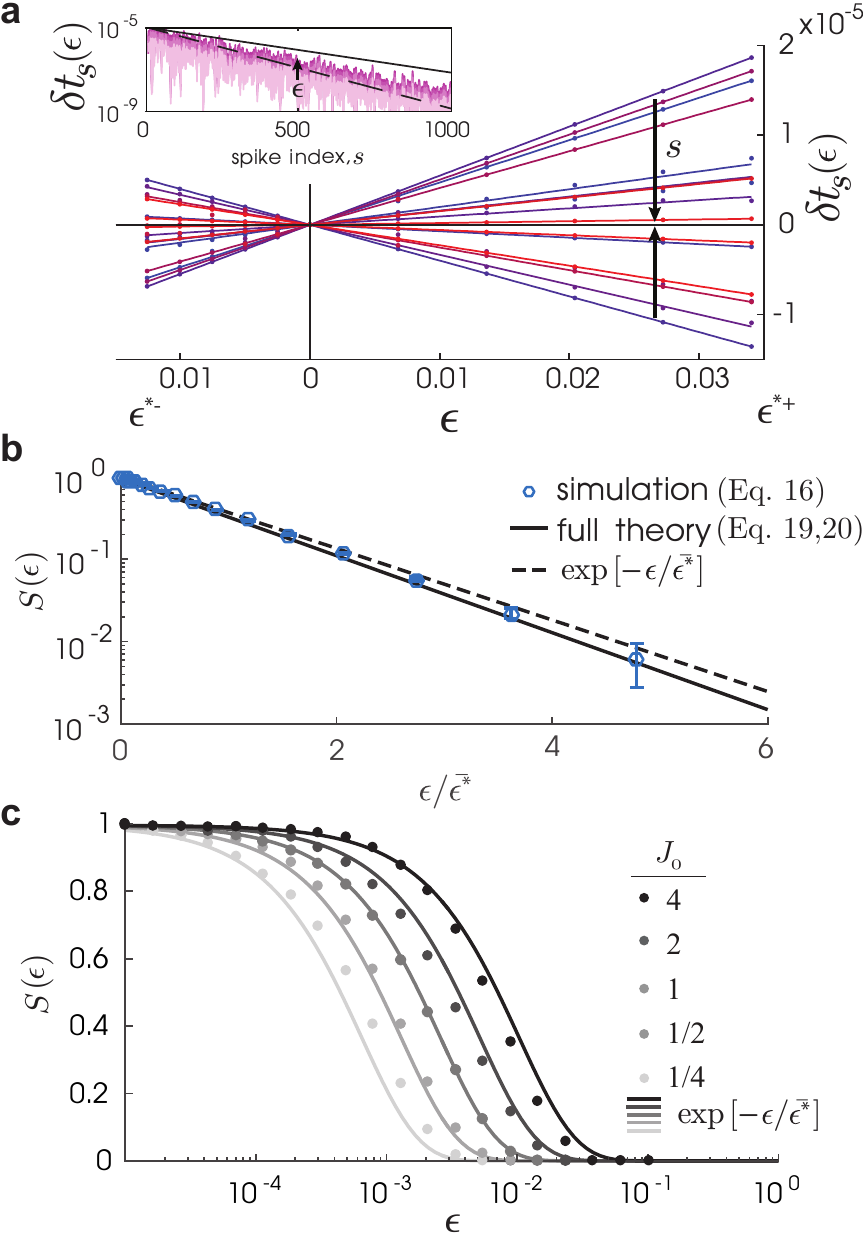}
\par\end{centering}
\caption{
The flux tube indicator function, once expressed with microstate variables and averaged, gives the survival probability to remain in the containing flux tube. 
\textbf{(a)} Spike-time deviations, $\delta t_{s}\left(\epsilon\right)$ (dots), as a function of perturbation strength up to the positive and negative critical strength, $\epsilon^{*-}$ and $\epsilon^{*+}$, respectively, for $s=1,\dots,15$ (colors) with their linear approximation (lines) given by equation \eqref{eq:smallepsapprox}.
Inset: $\delta t_{s}\left(\epsilon\right)$ as a function of $s$ (shown for $\epsilon=0.2\epsilon^{*\pm},0.4\epsilon^{*\pm},0.6\epsilon^{*\pm},0.8\epsilon^{*\pm}$) decays exponentially at a rate near the maximum and mean Lyapunov exponent, $\lambda_{\mathrm{max}}$ (black line) and $\lambda_{\mathrm{mean}}$ (black-dashed line) respectively \citep{Monteforte2012}. 
\textbf{(b)} The survival probability function $S\left(\epsilon\right)$ from simulations (dots, equation \eqref{eq:fracrestore}; bars are standard error), theory (line, equations \eqref{eq:microdef},\eqref{eq:Ss}), and the simplified theory at large $K$, $\exp\left[-\epsilon/\overbar{\epsilon^{*}}\right]$ (dotted line, equation \eqref{eq:final Fr estimate-1}), where $\overbar{\epsilon^{*}}=\left(\sqrt{KN}\bar{\nu}\tau/J_{0}\right)^{-1}$.
\textbf{(c)} $S\left(\epsilon\right)$ from simulations (dots) and $\exp\left[-\epsilon/\overbar{\epsilon^{*}}\right]$ (lines) for $J_{0}=2^{n},n=-2,-1,0,1,2$.
(Same parameters as Fig. \ref{fig:The-known-LIF} except $N=10^{4},\;K=10^{3}$.)
}
\label{fig:flinearization-1}
\end{figure}

As $\rho_{T}$ factorizes, so does $S\left(\epsilon\right)$, \vspace{5pt}
\begin{align}
S\left(\epsilon\right) & =\lim_{T\to\infty}\left[\prod_{s=1}^{M}S_{s}\left(\epsilon\right)\right]_{P(M)}=\prod_{s=1}^{\infty}S_{s}\left(\epsilon\right)\;,\label{eq:microdef}
\end{align}
where $S_{s}\left(\epsilon\right)$ is the probability that a perturbation of strength $\epsilon$ does not lead to a collision event involving the $s^{\mathrm{th}}$ spike. With the above simplifications, \vspace{5pt}
\begin{align}
S_{s}(\epsilon) & =\left[\Theta\left(\Delta t-C\Delta a_{s}\epsilon\right)^{A_{mn}}\right]_{\rho_{t}\left(\Delta t\right)\rho_{a}\left(\Delta a_{s}\right)P_{A_{mn}}\left(A_{mn}\right)}\;.\label{eq:Ss}
\end{align}
Evaluating equation \eqref{eq:Ss} (see Supplementary Notes for the derivation), we find \vspace{5pt}
\begin{align}
S_{s}(\epsilon) & =1+p\left(\mathrm{Erfcx}\left[x_{s}\right]-1\right)\;,\label{eq:Serf}
\end{align}
where $x_{s}=\frac{C}{\overbar{\Delta t}} e^{\frac{\lambda_{\mathrm{inh}}}{N}s}\epsilon\leq\epsilon/\sqrt{p}$, and $\mathrm{Erfcx}\left[x\right]=e^{x^{2}}\left(1-\frac{2}{\sqrt{\pi}}\int_{0}^{x}e^{-y^{2}}\mathrm{d}y\right)$ is the scaled complementary error function. 
$\mathrm{Erfcx}\left[x_{s}\right]-1\approx-x_{s}$ for $\epsilon/\sqrt{p}\ll1$, so that finally\vspace{5pt}
\begin{align*}
S\left(\epsilon\right)\approx & \prod_{s=1}^{\infty}\left(1-C\lambda_{\mathrm{sus}}e^{\frac{\lambda_{\mathrm{inh}}}{N}s}\epsilon\right)\;,
\end{align*}
where we have identified $\lambda_{\mathrm{sus}}=\frac{p}{\overbar{\Delta t}}$.
Employing the logarithm and $C\lambda_{\mathrm{sus}}\epsilon\propto\sqrt{p}\epsilon\ll1$,\vspace{5pt}
\begin{eqnarray}
S\left(\epsilon\right) & \approx & \exp\left[\frac{\lambda_{\mathrm{sus}}}{\frac{\lambda_{\mathrm{inh}}}{N}}C\epsilon\right]=\exp\left[-\frac{\epsilon}{\overbar{\epsilon^{*}}}\right]\label{eq:final Fr estimate-1}
\end{eqnarray}
with \vspace{5pt}
\begin{align}
\overbar{\epsilon^{*}} & =\frac{1}{C}\frac{\left|\frac{\lambda_{\mathrm{inh}}}{N}\right|}{\lambda_{\mathrm{sus}}} 
  =\frac{\sqrt{N}}{T_{\mathrm{free}}}\frac{pd}{p/\overbar{\Delta t}}=\sqrt{N}\overbar{\Delta t}\frac{d}{T_{\mathrm{free}}}\nonumber \\
 & =\sqrt{N}\overbar{\Delta t}\frac{\left|J\right|/I_{\mathrm{Ext}}}{\tau/I_{\mathrm{Ext}}}=\sqrt{N}\overbar{\Delta t}\frac{\left|J\right|}{\tau}=\frac{\sqrt{N}}{\tau}\frac{\left|-J_{0}/\sqrt{K}\right|}{N\bar{\nu}}\label{eq:simpleexpr}\\
\overbar{\epsilon^{*}} & =\frac{J_{0}}{\sqrt{KN}\bar{\nu}\tau}\;,\label{eq:final eps}
\end{align}
where we have used equations \eqref{eq:large K Tfree =000026 d} and \eqref{eq:dapprox} in the second line and note the cancellation of $p$ and $I_{\mathrm{Ext}}$.
Equation \eqref{eq:final Fr estimate-1} shows for $1\ll K\ll N$ that the basin diameter, $\epsilon^{*}$, is exponentially distributed and so completely determined by its characteristic scale, $\overbar{\epsilon^{*}}$ (equation \eqref{eq:final eps}), that is smaller for larger network size, higher average in-degree, higher population activity, and larger membrane time constant, $\tau$. 
$\overbar{\epsilon^{*}}$ grows, however, with the synaptic coupling strength, $J_{0}$. In Fig. \ref{fig:flinearization-1}b, we see quantitative agreement in simulations between the definition of $\hat{S}\left(\epsilon\right)$ (equation \eqref{eq:fracrestore} using the definition of $\epsilon^{*}$, equation \eqref{eq:epscritdef}) and its approximate microstate parametrization (equations \eqref{eq:microdef}, \eqref{eq:Ss}). 
These also confirm the exponential form of our reduced expression (equations \eqref{eq:final Fr estimate-1}, \eqref{eq:final eps}) and a scaling dependence on $J_{0}$ (Fig. \ref{fig:flinearization-1}c).
The latter holds until $J$ is no longer of size $\mathcal{O}\left(1/\sqrt{K}\right)$.
The other scalings were reported in Ref.\citep{Monteforte2012}. 
A derivation of only the characteristic scaling of $\overbar{\epsilon^{*}}$, but not depending on the Poisson spiking assumption, is given in the Supplemental Notes.

\section*{The geometry of phase space partitioning}
\begin{figure}[ht!]
\includegraphics{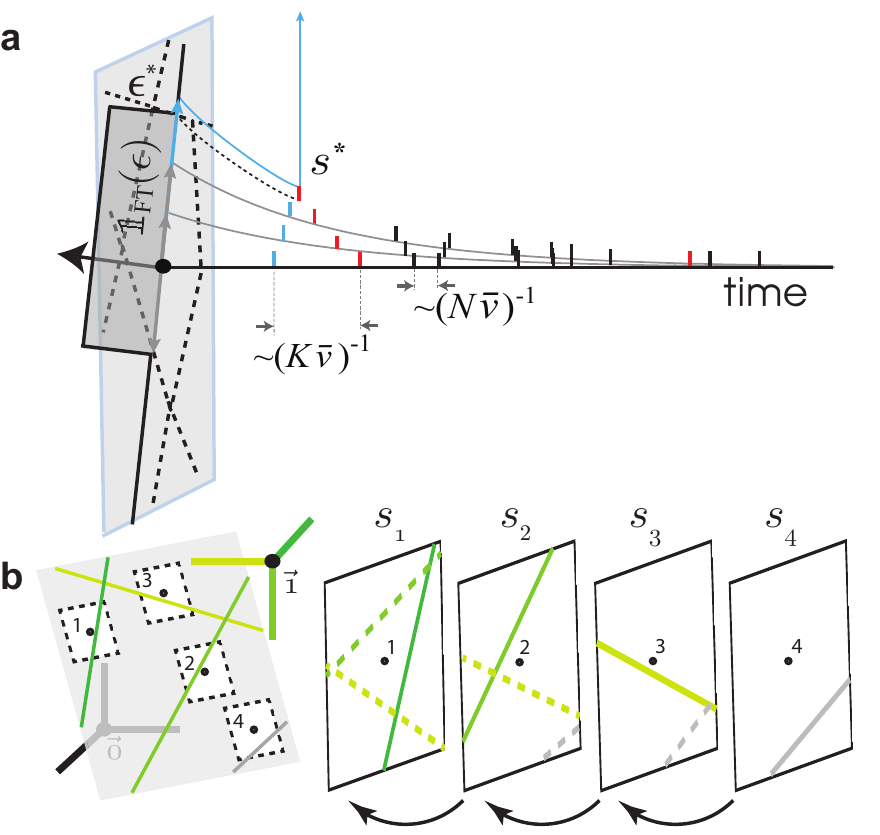}
\centering{}
\caption{
Flux tube boundaries are the pre-images of future susceptible spike collisions.
\textbf{ (a)} A folded phase space representation of a susceptible spike collision. 
Spikes (ticks) occur at a rate $N\bar{\nu}$ in the unperturbed trajectory (black line). 
For an example spike (blue tick), its susceptible spike partners (red ticks) occur at lower rate $K\bar{\nu}$. 
Small perturbations (gray arrows) lead to trajectories (gray) exhibiting spike time deviations that decay over time (tick alignment). 
A larger perturbation (example just beyond the critical perturbation strength: blue arrow) can push a spike and one of its susceptible partner spikes in the subsequent trajectory (blue and red, respectively) to collide, generating a divergence event at spike $s^{*}$. 
The indicator function, $\mathbb{1}_{\mathrm{FT}}\left(\epsilon\right)$, has support (dark gray) only over the local tube. 
\textbf{(b) }Constructing the local flux-tube partition in the non-folded phase space. 
\emph{Left}: Susceptible spikes are represented by susceptible edges (thick green lines) of the unit hypercube having $\vec{1}=(1,\dots,1)$ (black dot) as an endpoint. 
An intrinsic random partition (thin green lines) is generated by projecting these edges onto the hyper-plane orthogonal to $\vec{1}$  (light gray).
A given trajectory (labeled sequence of small dots) and its local neighborhood (within black dashed lines) is shown. 
\emph{Right}: The flux tube partition for this trajectory at a given spike (here $s_{1}$) is obtained from back-iterating the intrinsic partition from all future spikes (here only partitions from $s_{2}$, $s_{3}$, and $s_{4}$ are back-iterated; dashed lines). 
The partition at sufficiently distant future spikes (here the gray edge at $s_{4}$) will no longer refine the partition in the local neighborhood at $s_{1}$, since the expansive backwards dynamics maps the projected edges outside the neighborhood. 
A concrete example obtained from simulations is presented in the Supplemental Notes.}
\label{fig:Flux-tube-summary}
\end{figure}

Figure \ref{fig:Flux-tube-summary} presents the phase space organization of these spiking circuits as we have revealed it, replacing the caricature of Fig. \ref{fig:The-known-LIF}c. For a perturbation made to a stable trajectory, the geometry of the determining collision event is shown in Figure \ref{fig:Flux-tube-summary}a, in a folded representation. 
The pre-images of this event determine the flux tube boundary back to the perturbation. 
Our results also provide a global, i.e. non-folded geometry of the partitioning (Fig. \ref{fig:Flux-tube-summary}b(left)). 
Susceptible spike collisions are edges of the $N$-dimensional unit hypercube of phases where the corresponding voltages of two connected neurons both approach threshold. 
The Poincare section obtained by projecting the dynamics orthogonal to the trajectory (since no motion exists orthogonal to this subspace) then reveals the intrinsic partition. 
Here, the polygon basin boundaries arise as the pre-images of the projections of susceptible edges lying nearby the trajectory at future spike times (Fig. \ref{fig:Flux-tube-summary}b(right)).

\begin{discussion}

We have developed a theory of phase space partitioning in spiking neural circuits, exemplified using the phenomenon of flux tubes.
Importantly, the approach yields the dependence on various control parameters. 
We find the flux tube diameter contracts with the rate of volume contraction per neuron, $\lambda_{\mathrm{inh}}/N=(N\bar{\nu}\tau)^{-1}$, due to the inhibition received across the post-synaptic subspace of each spike.
This contraction is punctuated, however, by collision events between susceptible spikes, \emph{i.e.} those from pairs of connected neurons, occurring at rate  $\lambda_{\mathrm{sus}}=(K\bar{\nu})^{-1}$ and across which the basin volume expands out to a pre-image of the next collision event. For some neighboring tube, this collision event sets off a cascade of such events with exponential rate, $\lambda_{\mathrm{sus}}$ that is responsible for their mutual divergence.
Using these collision events to identify the spiking trajectories lying on flux tube boundaries, we were able calculate the size distribution of these basins. The average size is controlled by the ratio of these two exponential rates. Leaving out a factor converting shifts in spike time to shifts in state,\vspace{5pt}
\begin{align*}
\overbar{\epsilon^{*}} & \propto\frac{\left|J\right|}{N\bar{\nu}}\equiv\frac{\text{stabilizing\;inhibitory\;coupling\;strength}}{\text{destabilizing\;rate\ of\;spikes}}\;.
\end{align*}
The final scaling, $\overbar{\epsilon^{*}}= J_{0}/\left(\sqrt{KN}\bar{\nu}\tau\right)$, thus combines the contraction from the single neuron dynamics responsible for the dissipative dynamics, with the overall rate of spikes, which appears since each spike can be involved in a destabilizing collision event. 
Both contracting and expanding rates scale with the probability of connection, $p$, so we intuitively expect $p$ to appear in $\overbar{\epsilon^{*}}$ only implicitly through $J$ and, reassuringly, $p$ indeed cancels out.

Our framework motivates a variety of extensions. 
Our calculations can be performed for different disordered connectivity ensembles (e.g. correlated entries from annealed dilution processes \citep{Bouten1990} and structured second-order statistics \citep{Zhao2011}), different activity regimes (e.g. non-Markovian spike interval processes \citep{Schwalger2015}), and different single neuron models (e.g. any threshold neuron with known phase response curve). 
We have applied the theory to an instability caused by abrupt changes in spike time due to an inhibitory input near voltage threshold, a scenario that can also be analyzed in neuron models with smooth thresholds (e.g. the rapid theta-neuron \citep{Monteforte2011} that has the LIF neuron as a limit). The theory may also apply to other, as yet unknown instabilities involving spike collision events. Finally, while the linear stability of the dynamics precludes finite, asymptotic (Kolmogorov-Sinai) entropy production, the partition refinement picture we provide in Fig. \ref{fig:Flux-tube-summary}b suggests a transient production of information about the perturbation on timescales of the order of the divergence event time, $t^{*}$. 
Making this connection to ergodic theory more precise is an interesting direction for future research.

Applying our approach in a relatively idealized context allowed for a tractable assessment of phase space organization. 
Despite its simplicity, however, the LIF neuron accurately captures many properties of cortical neurons, such as their dynamic response \citep{Brette2015}. 
We have also neglected heterogeneity in many properties. For instance, in contrast to the locally stable regime studied here, mixed networks of excitatory and inhibitory neurons can instead be conventionally chaotic \citep{Monteforte2010}. 
This chaos can nevertheless be suppressed in the ubiquitous presence of fluctuating external drive \citep{Molgedey1992,Goedeke2016} or with spatially-structured connectivity \citep{Rosenbaum2014}, suggesting a generality to locally stable dynamics and phase space partitioning in neural computation. 
Our approach, in particular the way we have quantified the ensemble of perturbed spiking trajectories, can inform formulations of local stability in these more elaborate contexts. 
Of particular interest are extensions where a macroscopic fraction of tubes remain large enough to realize encoding schemes tolerant of intrinsic and stimulus noise. 
For example, extensions to random dynamical systems \citep{Lajoie2014,arnold2013random} could provide theoretical control over spiking dynamic variants of rate network-based learning schemes to generate stable, input-specific trajectories \citep{Laje2013}. 

Recent advances in experimental neuroscience have allowed for probes of the finite-size stability properties of cortical circuit dynamics call for \emph{in vivo}. 
For example, simultaneous intra- and extra-cellular recordings in the whisker motion-sensing system of the rat reveal that the addition of a single spike makes a measurable impact on the underlying spiking dynamics of the local cortical area \citep{London2010a}. 
Indeed, rats can be trained to detect perturbations to single spikes emitted in this area \citep{Houweling2008}. Representative toy theories, such as the one we provide, can guide this work by highlighting the features of spiking neural circuits that contribute to phase space partitioning. The combined effort promises to elucidate the dynamical substrate for neural computation at the level at which the neuronal interactions actually operate.

\end{discussion}

\section*{Acknowledgements}
M.P.T. would like to acknowledge discussions with Michael Monteforte, Sven Jahnke and Rainer Engelken. This work was supported by BMBF (01GQ07113, 01GQ0811, 01GQ0922, 01GQ1005B), GIF (906-17.1/2006), DFG (SFB 889), VW-Stiftung (ZN2632), and the Max Planck Society. 

\section*{Author Contributions}
M.P.T. conceived the project, developed the concepts, and wrote the manuscript. F.W. supervised the project, discussed the results and edited the manuscript. 

\section*{Additional Information}
The authors declare no competing financial interests.

\newpage

\newpage

\newpage

\newpage

\newpage

\newpage

\begin{thebibliography}{37}
\providecommand{\natexlab}[1]{#1}
\providecommand{\url}[1]{\texttt{#1}}
\expandafter\ifx\csname urlstyle\endcsname\relax
  \providecommand{\doi}[1]{doi: #1}\else
  \providecommand{\doi}{doi: \begingroup \urlstyle{rm}\Url}\fi

\bibitem[Vogels et~al.(2011)Vogels, Sprekeler, Zenke, Clopath, and
  Gerstner]{Vogels2011}
Vogels, T.~P., Sprekeler, H., Zenke, F., Clopath, C., and Gerstner, W.
\newblock {Inhibitory plasticity balances excitation and inhibition in sensory
  pathways and memory networks.}
\newblock \emph{Science},  {\bf 334} , 1569--73, 2011.

\bibitem[Wilson and Cowan(1972)]{Wilson1972}
Wilson, H.~R. and Cowan, J.~D.
\newblock {Excitatory and inhibitory interactions in localized populations of
  model neurons.}
\newblock \emph{Biophysical journal},  {\bf 12} , 1--24, 1972.

\bibitem[Sompolinsky et~al.(1988)Sompolinsky, Crisanti, and
  Sommers]{Sompolinsky1988}
Sompolinsky, H., Crisanti, A., and Sommers, H.~J.
\newblock {Chaos in random neural networks}.
\newblock \emph{Physical Review Letters},  {\bf 61} , 259--262, 1988.

\bibitem[Kadmon and Sompolinsky(2015)]{Kadmon2015}
Kadmon, J. and Sompolinsky, H.
\newblock {Transition to chaos in random neuronal networks}.
\newblock \emph{Physical Review X},  {\bf 5} , 1--28, 2015.

\bibitem[Hopfield(1982)]{Hopfield1982}
Hopfield, J.~J.
\newblock {Neural Networks and Physical Systems with Emergent Collective
  Computational Abilities}.
\newblock \emph{Proceedings of the National Academy of Sciences},  {\bf 79} ,
  2554--2558, 1982.

\bibitem[Sussillo and Abbott(2009)]{Sussillo2009}
Sussillo, D. and Abbott, L.~F.
\newblock {Generating Coherent Patterns of Activity from Chaotic Neural
  Networks}.
\newblock \emph{Neuron},  {\bf 63} , 544--557, 2009.

\bibitem[Laje and Buonomano(2013)]{Laje2013}
Laje, R. and Buonomano, D.~V.
\newblock {Robust timing and motor patterns by taming chaos in recurrent neural
  networks}.
\newblock \emph{Nature Neuroscience},  {\bf 16} , 925--933, 2013.

\bibitem[Brunel(2016)]{Brunel2016}
Brunel, N.
\newblock {Is cortical connectivity optimized for storing information?}
\newblock \emph{Nature Neuroscience},  {\bf 19} , 749--755, 2016.

\bibitem[LeCun et~al.(2015)LeCun, Bengio, and Hinton]{LeCun2015}
LeCun, Y., Bengio, Y., and Hinton, G.
\newblock {Deep learning}.
\newblock \emph{Nature},  {\bf 521} , 436--444, 2015.

\bibitem[Gardner(1988)]{Gardner1999}
Gardner, E.
\newblock {Optimal basins of attraction in randomly sparse neural network
  models}.
\newblock \emph{Journal of Physics A: Mathematical and General},  {\bf 22} ,
  1969--1974, 1988.

\bibitem[Harish and Hansel(2015)]{Harish2015}
Harish, O. and Hansel, D.
\newblock {Asynchronous Rate Chaos in Spiking Neuronal Circuits}.
\newblock \emph{PLoS Computational Biology},  {\bf 11} , e1004266, 2015.

\bibitem[Mastrogiuseppe and Ostojic(2017)]{Mastrogiuseppe2017}
Mastrogiuseppe, F. and Ostojic, S.
\newblock {Intrinsically-generated fluctuating activity in
  excitatory-inhibitory networks}.
\newblock \emph{PLOS Computational Biology},  {\bf 13} , 1--40, 2017.

\bibitem[Engelken et~al.(2016)Engelken, Farkhooi, Hansel, van Vreeswijk, and
  Wolf]{Engelken2016}
Engelken, R., Farkhooi, F., Hansel, D., van Vreeswijk, C., and Wolf, F.
\newblock {A reanalysis of “Two types of asynchronous activity in networks of
  excitatory and inhibitory spiking neurons”}.
\newblock \emph{F1000Research},  {\bf 5} , 2043, 2016.

\bibitem[Roxin et~al.(2011)Roxin, Brunel, Hansel, Mongillo, and {Van
  Vreeswijk}]{Roxin2011}
Roxin, A., Brunel, N., Hansel, D., Mongillo, G., and {Van Vreeswijk}, C.
\newblock {On the distribution of firing rates in networks of cortical
  neurons.}
\newblock \emph{J Neurosci},  {\bf 31} , 16217--26, 2011.

\bibitem[Memmesheimer et~al.(2014)Memmesheimer, Rubin, {\"{O}}lveczky, and
  Sompolinsky]{Memmesheimer2014}
Memmesheimer, R.~M., Rubin, R., {\"{O}}lveczky, B., and Sompolinsky, H.
\newblock {Learning Precisely Timed Spikes}.
\newblock \emph{Neuron},  {\bf 82} , 925--938, 2014.

\bibitem[Jahnke et~al.(2008)Jahnke, Memmesheimer, and Timme]{Jahnke2008}
Jahnke, S., Memmesheimer, R.-M., and Timme, M.
\newblock {Stable Irregular Dynamics in Complex Neural Networks}.
\newblock \emph{Physical Review Letters},  {\bf 100} , 2--5, 2008.

\bibitem[Monteforte and Wolf(2012)]{Monteforte2012}
Monteforte, M. and Wolf, F.
\newblock {Dynamic flux tubes form reservoirs of stability in neuronal
  circuits}.
\newblock  , {\bf 2}, 041007, 2012.

\bibitem[van Vreeswijk and Sompolinsky(1996)]{VanVreeswijk1996}
van Vreeswijk, C. and Sompolinsky, H.
\newblock {Chaos in neuronal networks with balanced excitatory and inhibitory
  activity.}
\newblock \emph{Science (New York, N.Y.)},  {\bf 274} , 1724--6, 1996.

\bibitem[Brunel and Hakim(1999)]{Brunel1999}
Brunel, N. and Hakim, V.
\newblock {Fast global oscillations in networks of integrate-and-fire neurons
  with low firing rates.}
\newblock \emph{Neural Computation},  {\bf 11} , 1621--71, 1999.

\bibitem[Renart et~al.(2010)Renart, de~la Rocha, Bartho, Hollender, Parga,
  Reyes, and Harris]{Renart2010}
Renart, A., de~la Rocha, J., Bartho, P., Hollender, L., Parga, N., Reyes, A.,
  and Harris, K.~D.
\newblock {The asynchronous state in cortical circuits.}
\newblock \emph{Science},  {\bf 327} , 587--590, 2010.

\bibitem[Barral and {D Reyes}(2016)]{Barral2016}
Barral, J. and {D Reyes}, A.
\newblock {Synaptic scaling rule preserves excitatory–inhibitory balance and
  salient neuronal network dynamics}.
\newblock \emph{Nature Neuroscience},  {\bf 19} , 1690--1696, 2016.

\bibitem[Jin(2002)]{Jin2002}
Jin, D.
\newblock {Fast Convergence of Spike Sequences to Periodic Patterns in
  Recurrent Networks}.
\newblock \emph{Physical Review Letters},  {\bf 89} , 1--4, 2002.

\bibitem[Tuckwell(1988)]{Tuckwell1988}
Tuckwell, H.
\newblock \emph{{Introduction to Theoretical Neurobiology vols. 1 and 2}}.
\newblock Cambridge University Press, 1988.

\bibitem[Lindner(2006)]{Lindner2006}
Lindner, B.
\newblock {Superposition of many independent spike trains is generally not a
  Poisson process}.
\newblock \emph{Physical Review E},  {\bf 73} , 1--4, 2006.

\bibitem[Boutent et~al.(1990)Boutent, Engels, Komodat, and
  Serneelst]{Bouten1990}
Boutent, M., Engels, A., Komodat, A., and Serneelst, R.
\newblock {Quenched versus annealed dilution in neural networks}.
\newblock \emph{J. Phys. A: Math. Gen},  {\bf 23} , 4643--4657, 1990.

\bibitem[Zhao et~al.(2011)Zhao, Beverlin, Netoff, and Nykamp]{Zhao2011}
Zhao, L., Beverlin, B., Netoff, T., and Nykamp, D.~Q.
\newblock {Synchronization from second order network connectivity statistics.}
\newblock \emph{Frontiers in computational neuroscience},  {\bf 5} , 28, 2011.

\bibitem[Schwalger et~al.(2015)Schwalger, Droste, and Lindner]{Schwalger2015}
Schwalger, T., Droste, F., and Lindner, B.
\newblock {Statistical structure of neural spiking under non-poissonian or
  other non-white stimulation}.
\newblock \emph{Journal of Computational Neuroscience},  {\bf 39} , 29--51,
  2015.

\bibitem[Monteforte(2011)]{Monteforte2011}
Monteforte, M.
\newblock \emph{{Chaotic Dynamics in Networks of Spiking Neurons in the
  Balanced State}}.
\newblock PhD thesis, 2011.

\bibitem[Brette(2015)]{Brette2015}
Brette, R.
\newblock {What Is the Most Realistic Single-Compartment Model of Spike
  Initiation?}
\newblock \emph{PLOS Computational Biology},  {\bf 11} , e1004114, 2015.

\bibitem[Monteforte and Wolf(2010)]{Monteforte2010}
Monteforte, M. and Wolf, F.
\newblock {Dynamical Entropy Production in Spiking Neuron Networks in the
  Balanced State}.
\newblock \emph{Phys. Rev. Lett.},  {\bf 105(26)} , 1--4, 2010.

\bibitem[Molgedey et~al.(1992)Molgedey, Schuchhardt, and
  Schuster]{Molgedey1992}
Molgedey, L., Schuchhardt, J., and Schuster, H.~G.
\newblock {Suppressing chaos in neural networks by noise}.
\newblock \emph{Physical Review Letters},  {\bf 69} , 3717--3719, 1992.

\bibitem[Goedeke et~al.(2016)Goedeke, Schuecker, and Helias]{Goedeke2016}
Goedeke, S., Schuecker, J., and Helias, M.
\newblock {Noise dynamically suppresses chaos in neural networks}.
\newblock  , pages 1--5, 2016.

\bibitem[Rosenbaum and Doiron(2014)]{Rosenbaum2014}
Rosenbaum, R. and Doiron, B.
\newblock {Balanced networks of spiking neurons with spatially dependent
  recurrent connections}.
\newblock \emph{Physical Review X},  {\bf 4} , 1--9, 2014.

\bibitem[Lajoie et~al.(2014)Lajoie, Thivierge, and Shea-Brown]{Lajoie2014}
Lajoie, G., Thivierge, J., and Shea-Brown, E.
\newblock {Structured chaos shapes spike-response noise entropy in balanced
  neural networks}.
\newblock \emph{Frontiers in computational neuroscience},  {\bf 8} , 2014.

\bibitem[Arnold(1998)]{arnold2013random}
Arnold, L.
\newblock \emph{{Random Dynamical Systems}}.
\newblock Springer, 1998.

\bibitem[London et~al.(2010)London, Roth, Beeren, H{\"{a}}usser, and
  Latham]{London2010a}
London, M., Roth, A., Beeren, L., H{\"{a}}usser, M., and Latham, P.~E.
\newblock {Sensitivity to perturbations in vivo implies high noise and suggests
  rate coding in cortex.}
\newblock \emph{Nature},  {\bf 466} , 123--127, 2010.

\bibitem[Houweling and Brecht(2008)]{Houweling2008}
Houweling, A.~R. and Brecht, M.
\newblock {Behavioural report of single neuron stimulation in somatosensory
  cortex.}
\newblock \emph{Nature},  {\bf 451} , 65--8, 2008.

\end{thebibliography}
\end{document}


\sloppy
 \pagenumbering{gobble}

\maketitle

\begin{affiliations}
\item Max Planck Institute for Dynamics and Self-Organization, G{\"o}ttingen, Germany
\item Bernstein Center for Computational Neuroscience, G{\"o}ttingen, Germany
\item Laboratoire de Physique Th{\'e}orique, ENS-PSL Research University, Paris, France
\item Kavli Institute for Theoretical Physics, University of California Santa Barbara, Santa Barbara, USA
\end{affiliations}

\tableofcontents

\clearpage
\pagenumbering{arabic}  
\linenumbers
\section{Supplementary Methods}
\subsection{Event-based simulations of inhibitory LIF networks}

As described in the Methods section of the main text, we analyze the
inhibitory LIF network dynamics. A true phase representation is defined
on a circular domain, e.g. $\left[0,1\right]^{N}$ with $0$ and $1$
identified. The phase dynamics we analyze is a \emph{pseudo}phase
representation, $\vec{\phi}(t)\in\left(-\infty,1\right]^{N}$(\emph{pseudo}
since the phase can be knocked, to a negative value by an inhibitory
input near the reset, $V\approx V_{R}$). We hereon drop the \emph{pseudo}
from the terminology for clarity. The phase representation dynamics
is given by: 
\begin{equation}
\dot{\phi}_{n}\left(t\right)=T_{\mathrm{free}}^{-1}+\sum_{s}A_{nn_{s}}\delta(t-t_{s})Z(\phi_{n}\left(t_{s}\right))
\end{equation}
with constant phase velocity, $T_{\mathrm{free}}^{-1}$, the phase
response curve, $Z(\phi)$, and a spike-reset rule: when $\phi_{n}=1$,
$\phi_{n}$ is reset to $0$. Note that in the large-$K$ limit the
phase and the voltage representation converge onto one another (see Ref. 15).

Event-based simulations were implemented by iterations of a map from
just after one spike in the network, $t_{s}$, to just after the next,
$t_{s+1}$, where $s$ is the index of the network spike sequence.
The next spike time, $t_{s+1}$, and next spiking neuron in the sequence
are obtained simply in the phase representation via 
\begin{eqnarray*}
t_{s+1} & = & t_{s}+\min_{n\in\{1,\dots,N\}}\left(1-\phi_{n}(t_{s})\right)T_{\mathrm{free}}\;,\\
n_{s+1} & = & \mbox{argmin}_{n\in\left\{ 1,\dots,N\right\} }\left(1-\phi_{n}(t_{s})\right)T_{\mathrm{free\;,}}
\end{eqnarray*}
respectively. An iteration consists of evolving the network phases
to this next spike time, $t_{s+1}$, applying the pulse of size $Z(\phi_{m}\left(t_{s+1}\right))$
to the postsynaptic neurons, $\left\{ m|A_{mn_{s+1}}=1\right\} $,
and then resetting the phase of the spiking neuron, $n_{s+1}$.
For further details, as well as methods for computing the Lyapunov
spectrum for this network (see Ref. 15). In the remainder
of this section, we show how the large-$K$ expressions for $T_{\mathrm{free}}$,
and $Z'(\phi)$, equations (5) and (6) in main text, respectively, are obtained.

The period of firing in the absence of recurrent input, $T_{\mathrm{free}}$,
is obtained from the solution of the model (equations 1,2), 
\begin{equation}
V(t)=\sqrt{K}I_{0}-\left(\sqrt{K}I_{0}-V(0)\right)e^{-\frac{t}{\tau}}\;.\label{eq:voltsol}
\end{equation}
 with reference time , $t=0$, at which the initial condition $V(0)=V_{R}=-1$. The solution is used to
evolve the state to the threshold voltage, $V=V_{T}=0$, and then
inverted to obtain the elapsed time,
\begin{eqnarray*}
T_{\mathrm{free}} & = & \tau\ln\left[\frac{\sqrt{K}I_{0}-V_{R}}{\sqrt{K}I_{0}-V_{T}}\right]\\
 & = & \tau\ln\left[1+\frac{1}{\sqrt{K}I_{0}}\right]\\
T_{\mathrm{free}} & \overset{K\gg1}{\approx} & \frac{1}{KJ_{0}\bar{\nu}}\;,
\end{eqnarray*}
where we have used the balance equation, equation 3, in the large-$K$
limit to express the external drive, 
\begin{equation}
I_{0}\approx I_{bal}:=J_{0}\bar{\nu}\tau\;,\label{eq:Ibal}
\end{equation}
 in terms of the population firing rate, $\bar{\nu}$, and coupling
strength, $J_{0}$. The phase, $\phi$, over this period satisfies 
\begin{equation}
\phi=\frac{t}{T_{\mathrm{free}}}\;.\label{eq:phasedef}
\end{equation}
The phase response curve, $Z(\phi)$, is the state-dependent change
to phase, $\phi$, as a result of an input spike. Its calculation for the LIF
neuron model is made by mapping phase to voltage, applying the inhibitory
synaptic input and then mapping back to phase. This procedure provides
the phase transition curve, $\phi_{after\;spike}=PTC(\phi_{before\;spike})$,
from which $\phi$ is subtracted to obtain the phase response curve,
$Z(\phi)=PTC(\phi)-\phi$. The transformation from phase to voltage
is obtained from equation \ref{eq:voltsol} using the definition of phase,
equation \ref{eq:phasedef},
\begin{align*}
V(\phi) & =\sqrt{K}I_{0}-\left(\sqrt{K}I_{0}+1\right)e^{-\frac{T_{\mathrm{free}}}{\tau}\phi}\;.
\end{align*}
The inhibitory synaptic input, $-J_{0}/\sqrt{K}$, is added and then
this new voltage is mapped back to phase using equation 4 from the main
text,
\begin{eqnarray*}
\phi_{after\;spike} & = & \frac{\tau}{T_{\mathrm{free}}}\ln\left[\frac{\sqrt{K}I_{0}+1}{\sqrt{K}I_{0}-\left(\sqrt{K}I_{0}-\left(\sqrt{K}I_{0}+1\right)e^{-\frac{T_{\mathrm{free}}}{\tau}\phi_{before\;spike}}-J_{0}/\sqrt{K}\right)}\right]\\
\phi_{after\;spike} & =- & \frac{\tau}{T_{\mathrm{free}}}\ln\left[e^{-\frac{T_{\mathrm{free}}}{\tau}\phi_{before\;spike}}+\frac{J_{0}}{\sqrt{K}\left(\sqrt{K}I_{0}+1\right)}\right]
\end{eqnarray*}
The phase response curve is then
\begin{eqnarray*}
Z(\phi) & = & -\frac{\tau}{T_{\mathrm{free}}}\ln\left(e^{-\frac{T_{\mathrm{free}}}{\tau}\phi}+\frac{J_{0}}{\sqrt{K}(\sqrt{K}I_{0}+1)}\right)-\phi\;.
\end{eqnarray*}
The derivative of $Z(\phi)$ is:
\begin{align*}
Z'(\phi) & =\frac{e^{-\frac{T_{\mathrm{free}}}{\tau}\phi}}{e^{-\frac{T_{\mathrm{free}}}{\tau}\phi}+\frac{J_{0}}{\sqrt{K}(1+\sqrt{K}I_{0})}}-1\\
Z'(\phi) & =\frac{1}{1+\frac{J_{0}}{\sqrt{K}(1+\sqrt{K}I_{0})}\left(1+\frac{1}{\sqrt{K}I_{0}}\right)^{\phi}}-1\;.
\end{align*}
where we have used the definition of $T_{\mathrm{free}}$ to re-express
$e^{-\frac{T_{\mathrm{free}}}{\tau}\phi}$ as $\left(1+\frac{1}{\sqrt{K}I_{0}}\right)^{\phi}$. 

$Z'(\phi)$ becomes independent of the phase in the large-$K$ approximation:
\begin{eqnarray*}
Z'(\phi) & = & -\frac{J_{0}}{KI_{0}}+\frac{J_{0}(1-\phi)}{K^{3/2}I_{0}^{2}}+\mathcal{O}\left(K^{-2}\right)\\
 & = & -\frac{1}{K\bar{\nu}\tau}-\frac{1-\phi}{J_{0}K^{3/2}\nu^{2}\tau^{2}}+\mathcal{O}\left(K^{-2}\right)\\
Z'(\phi) & \overset{K\gg1}{\approx} & -\frac{1}{K\bar{\nu}\tau}\;,
\end{eqnarray*}
where in the second line we again re-express $I_{0}$ using $I_{bal}$
from equation \ref{eq:Ibal}. The phase independence of $Z'(\phi)$ arises
from the linearization of the spike time change with vanishing synaptic
strength, $J\propto1/\sqrt{K}$ for $K\gg1$.

\subsection{Estimating the critical perturbation strength\label{sec:Critical-perturbation-strengths}}

A random perturbation direction, $\vec{\xi}$, was obtained by sampling
$N-1$ times from a standard normal distribution, normalizing this
vector, and projecting it into the $N$-dimensional phase space such
that it was orthogonal to the phase velocity vector $\vec{\omega}=\left(T_{\mathrm{free}}^{-1},\dots,T_{\mathrm{free}}^{-1}\right)$.
Constrained to this hyper-plane, the perturbation alters only relative
spike time differences, i.e. there is no global shift in spike times.
For $\epsilon>0$, the critical perturbation size, $\epsilon^{*}$,
in that direction was obtained using a bisection method, in which
the initial estimate of $\epsilon^{*}$, $\epsilon_{0}^{*}=J_{0}/\left(\sqrt{KN}\bar{\nu}\tau\right)$
was lower-bounded by $\epsilon_{low}^{*}=10^{-4}\cdot\epsilon_{0}^{*}$,
and upper-bounded by $\epsilon_{up}^{*}=1$. The estimate $\epsilon_{0}$
was iteratively refined based on a divergence flag on the distance between the perturbed and unperturbed trajectories
at time $T$ after the perturbation:
\begin{eqnarray*}
\mathrm{If} & \;D_{T}(\epsilon_{i})>D_{thresh} & ,\\
 & \mathrm{then}\;\;\epsilon_{\mathrm{upp}}^{*}\leftarrow\epsilon_{i}^{*} & ;\\
\mathrm{else}\\
 & \epsilon_{\mathrm{low}}^{*}\leftarrow\epsilon_{i}^{*} & ,
\end{eqnarray*}
for iteration index $i$, where $D_{thresh}=0.01$ denotes the threshold
chosen to lie between the two well-separated modes of the end-distance
distribution. ($D_{t}$ eventually saturates due to the bounded phase-space
at the average distance, $\bar{D}$, between a pair of random trajectories,
and computed in see Ref. 15.) A bisection step was then
made, 
\begin{equation}
\epsilon_{i+1}^{*}=\frac{\epsilon_{\mathrm{upp}}^{*}+\epsilon_{\mathrm{low}}^{*}}{2}\;,
\end{equation}
to obtain the estimate of the next iteration. The procedure was repeated
until the differences in successive values of $\epsilon_{i}^{*}$
fell below a tolerance threshold of $10^{-8}$, and the final estimate
taken as $\epsilon^{*}$.

\subsection{Constructing the folded phase space representation}
Here, we describe the procedure used to construct the folded representation
of the phase space around the attracting trajectory shown in Fig.2b and he Supplemental Video.
Similar to Fig. 7 in Ref. 15, the same, random 2D projection
of the $(N-1)$-dimensional subspace orthogonal the trajectory was
applied at each iteration of the event map. This subspace remains unchanged by the evolution since in the phase
representation the trajectory is always parallel to the main diagonal
of the unit hyper cube. Then, a rectilinear grid of initial conditions
were generated in these planes. The network was simulated from each
initial condition and the corresponding grid of end-states stored.
A corresponding grid of the pairwise distances between end-states
of all adjacent initial conditions was computed. Distances falling
in the finite-distance mode of the resulting bi-modal end-state distance
distribution centered around the average distance, $\bar{D}$, were
used to identify adjacent initial conditions spanning a putative flux
tube boundary. A putative tube identity label was assigned to each
continuous region of corresponding initial conditions enclosed by these putative
boundaries in the grid. We occasionally observed single tubes segregated
into disjoint pieces in our 2D representation by the occlusion of another tube, consistent
with the layering of projections as proposed in Fig. 5b. For robustness
then, a round of amalgamation of tube identities was performed by
identifying as the same any two tubes whose centers of mass gave a
end-state distance which fell below a threshold of 0.01. Again, that the
modes were well separated made for unambiguous flagging.

This algorithm to compute a single cross-section was then repeated
at each spike of the network activity in a simulated time window to
obtain a set of successive cross sections orthogonal to and centered
on the stable trajectory. To present this data, a folded representation
is used in which these cross sections are placed contiguously so that
the center trajectory passes through them continuously. This gives
a 2+1D representation of the tube and its neighborhood along the stable
trajectory, oriented such that the line $(0,0,t)$ is horizontal with
time increasing to the right. The identity of the center tube is trivially
maintained across sections since the $(0,0)$-perturbation leaves
the stable trajectory unchanged. To keep track of the identities of
the surrounding tubes represented in the successive sections requires
an identity list passed forward and updated from section to section.
We constructed such a list by again comparing all pairwise end-state
distances of the center of masses of all cells of the previous and
current cross sections and identifying successive cells as coming
from the same tube if this distance fell below a threshold. Identities
were added when a current cell has no match in the previous section
corresponding to the event of a new tube entering the section. Identities
were removed when a cell in the previous section had no match in the
current section corresponding to the event of an existing tube leaving
the section. We then used this identity list to color the cells, using
an adaptive color assignment scheme in order to keep the range of
colors reasonably bounded. This scheme randomly assigned unused colors,
orphaned from tubes that had exited the section, to the cells of new
tubes that had entered the section.

\section{Supplementary Video}
\href{file:http://www.phys.ens.fr/~mptouzel/pdf/PuelmaTouzel_Partioning_suppvideo.avi}{Supplementary Video}. Caption: A video of the evolution of the local phase space shown in Fig. 2b of the main text. 
In a folded representation around a given attracting trajectory, set at the origin, the simulation demonstrates the relative dynamics of the partitioning of the local phase space. 
Note that time is slowed relative to real time by a factor of approximately 6 (see Fig. 2 and corresponding text in the manuscript for further details). 
Spiking activity on the unperturbed attracting trajectory is shown below for reference (different shading denotes different spiking neurons).

\section{Supplementary Notes}
\subsection{Collision Motifs}

\subsubsection{Confirmation of decorrelation event properties}

In this section, we determine from simulations of the dynamics that
(1) the perturbed trajectories begin to diverge where a difference
in the spike sequence appears; (2) this change is associated with
a vanishing interval; and (3) this interval is between susceptible
spikes, i.e spikes from a pair of neurons that exhibit one of the
three connected-pair motifs. 

Over perturbation directions, an ensemble of pairs of perturbed trajectories
were simulated using a perturbation strength just above, $\epsilon^{*+}$,
and just below, $\epsilon^{*-}$, the estimate obtained according
to the procedure described in Section \ref{sec:Critical-perturbation-strengths}
(notation: $x^{\pm}=\lim_{\delta\to0}x\pm\delta$). From the simulation
started at $\epsilon^{*^{+}}$, the decorrelation index, $s^{*}$,
was extracted as the index in the spike sequence at which a sustained
difference between the pair of sequences began. We denote elements
of the perturbed spiking neuron sequence and spike times as $n_{s}\left(\epsilon\right)$
and $t_{s}\left(\epsilon\right)$, respectively.

\begin{figure}
\centering
\includegraphics[scale=0.9]{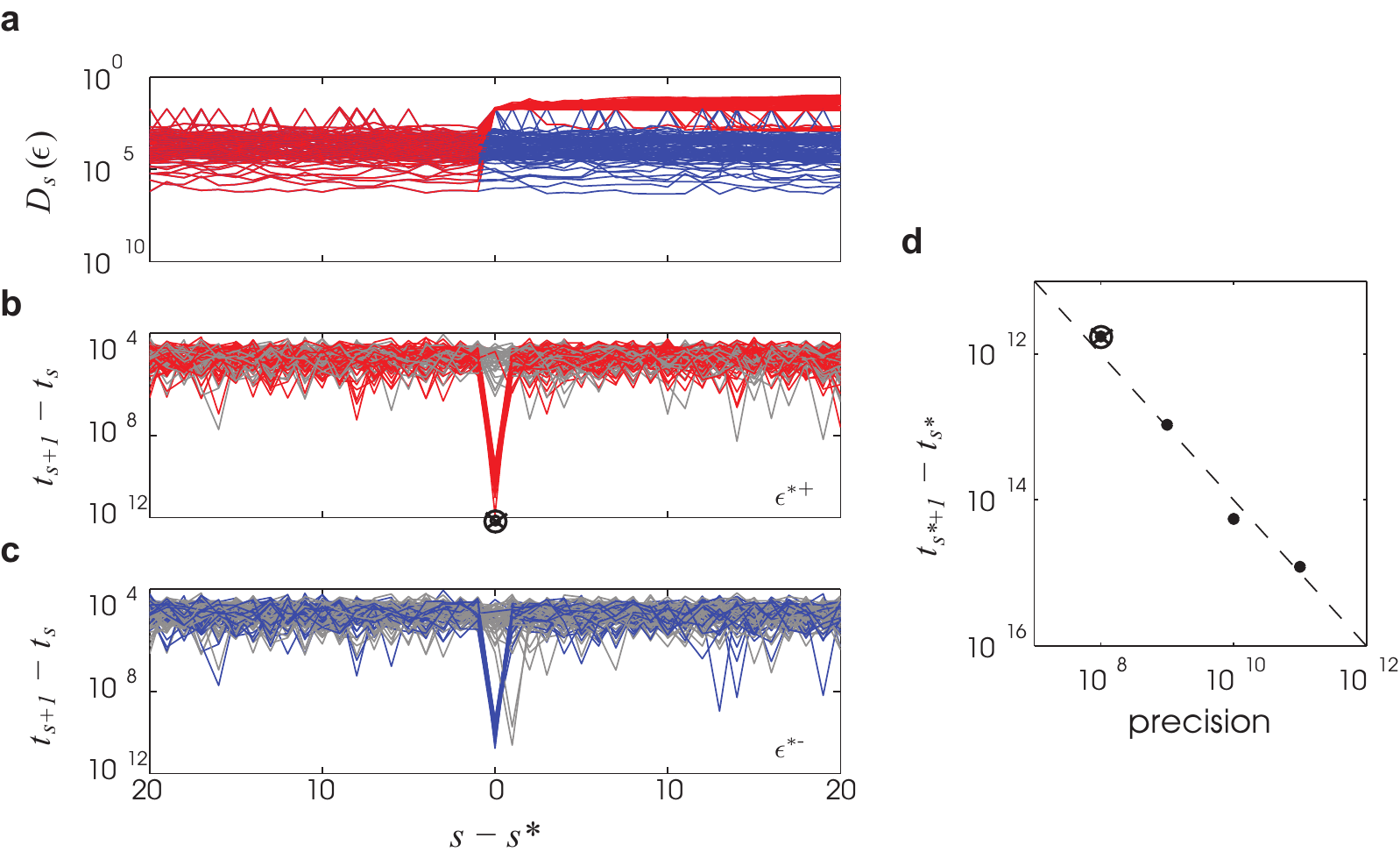}
\caption{Characteristics of divergence events. \textbf{(a)} A small window
of the distance time series aligned to index, $s^{*}$, at which the
decorrelation of the spike sequence begins. The supercritical perturbed (red) trajectory
started from $\epsilon^{*+}$ jumps up away from the subcritical perturbed (blue)
trajectory started at $\epsilon^{*-}$ at $s^{*}$ (note the logarithmic
scale on the ordinate). \textbf{(b)} and \textbf{(c)} show the spike
intervals, $t_{s+1}-t_{s}$, for the $\epsilon=\epsilon^{*+}$ and
$\epsilon=\epsilon^{*-}$ trajectories, respectively (colors as in
(a)). Realizations not exhibiting the $n_{s^{*-}}\rightarrow n_{s^{*-}+1}$
and $n_{s^{*-}}\leftarrow n_{s^{*-}+1}$, respectively, have been
grayed out. Note that those left colored have a significantly smaller
interval at index $s^{*}$.\textbf{ (d)} Coincidence of successive
spikes with increasing precision (decreasing tolerance) of the bisection
algorithm used to find $\epsilon^{*}$. Here, a shrinking interval
taken from a $\epsilon=\epsilon^{*+}$ realization has been used (see
the identified minimum in panel b. \label{fig:Distance-jump-and}}
\end{figure}

We first show that the sustained jump in distance begins at the decorrelation
index, $s^{*}$. We aligned by $s^{*}$ across trials the distances,
$D_{s}(\epsilon^{*+})$, to the unperturbed trajectory from the perturbed
trajectory started from $\epsilon^{*}$. The result, in Fig. \ref{fig:Distance-jump-and}a
shows the high correlation across trials. 

Next, in Fig. \ref{fig:Distance-jump-and}b,c we see that the spike
time interval, $t_{s^{*}+1}\left(\epsilon\right)-t_{s^{*}}\left(\epsilon\right)$
corresponding to $s^{*}$ before ($\epsilon=\epsilon^{*-}$) and after
($\epsilon=\epsilon^{*+}$) the collision event, respectively, vanishes
only when $A_{n_{s^{*}}n_{s^{*}+1}}=1$. In addition, $t_{s^{*}+1}\left(\epsilon\right)-t_{s^{*}}\left(\epsilon\right)$
scales inversely with the precision of the bisection algorithm used
to obtain $\epsilon^{*}$, demonstrating that the event is indeed
generated as two spikes become coincident, $t_{s^{*}+1}\left(\epsilon\right)\to t_{s^{*}}\left(\epsilon\right)$
 as $\epsilon\to\epsilon^*$ (see \ref{fig:Distance-jump-and}d).

\subsubsection{Spike crossing motifs}

In the main text we focused on the backward-connected motif. In this
section, we discuss the forward-connected and symmetrically-connected
motif. Across these motifs, under consideration is a situation where
an output spike time of a given neuron, $t_{out}$, is near in time
to an input spike time, $t_{in}$, that this neuron receives. When
the output spike is generated before the input spike, $t_{out}<t_{in}$
(the backward-connected motif), a collision can occur when a perturbation
leads to the vanishing of the interval between them, an example of
which is shown in Fig.3 in main text. If $t_{in}<t_{out}$ (the forward-connected
motif), however, the inhibition means that $t_{in}$ already delays $t_{out}$ for $\epsilon<\epsilon^*$ 
so that $t_{out}$ can occur no closer to $t_{in}$ than $\Delta t_{\mathrm{jump}}$
(see equation 9), for the same reason that $t_{out}$ undergoes a jump
forward in the backward-connected motif. Thus, a collision event occurs
in this motif when the perturbation brings $t_{in}$ and $t_{out}$
to within $\Delta t_{\mathrm{jump}}$ of each other. 

The two asymmetric motifs give collision scenarios that are identified
under a reversal of the direction of change in perturbation strength.
The forward and backward connected motif can be distinguished by whether
the collision event is approached by an input spike moving forward,
$\mbox{d}t_{in}/\mbox{d}\epsilon>0$, or backward, $\mbox{d}t_{in}/\mbox{d}\epsilon<0$,
over $t_{out}$ with $t_{out}$ as the reference time. In the forward-connected
motif, the interval vanishes, $t_{in}\to0^{+}$, for $\epsilon\to\epsilon^{*+}$,
\emph{i.e.} just after the collision. In the backward-connected motif,
the vanishing interval, $t_{in}\to0^{+}$, occurs as $\epsilon\to\epsilon^{*-}$,
\emph{i.e.} just before the collision. For either case, when on the
side of $\epsilon^{*}$ where the interval is vanishing, the input
spike comes after the output spike, $t_{in}>0$, in this reference
frame.

In each of these two asymmetric motifs, only one of the pair of spikes
undergoes a jump of size $\Delta t_{\mathrm{jump}}$. For the bidirectionally
connected motif, however, both spikes undergo a jump of size $\Delta t_{\mathrm{jump}}$
simultaneously, by which they exchange spike times, and so no vanishing
interval exists on either side of the flux tube boundary. A collision
event occurs in this motif with reduced relative frequency, $p$,
compared with the two asymmetric cases and so is negligible for sparse
networks, $p\ll1$.

The characteristics of an inhibitory event at threshold is a single
neuron property, dependent on the neuron model, and so can be investigated
for many neuron models. Since the LIF solution, equation \ref{eq:voltsol},
is invertible, one can explicitly solve for the time, $\Delta t_{\mathrm{jump}}$,
that the inhibitory event has delayed the spike. With initial condition,
$V(0)=V_{T}^{-}+J\approx J$, 
\begin{align}
V_{T} & =\sqrt{K}I_{0}-\left(\sqrt{K}I_{0}+J\right)e^{-\frac{\Delta t_{\mathrm{jump}}}{\tau}}\nonumber \\
\Delta t_{\mathrm{jump}} & =\tau\ln\left(1+\frac{J}{\sqrt{K}I_{0}}\right)\label{eq:spiketimeshift-1}
\end{align}
Using the balance equation, equation 3, we obtain $\Delta t_{\mathrm{jump}}\sim\tau\ln\left(1+\left(K\bar{\nu}\tau\right)^{-1}\right)\sim\left(K\bar{\nu}\right)^{-1}$,
for $K\gg1$, as stated in the main text. 

The inhibition prohibits susceptible spike pairs in the forward-connected
(and bidirectional) motif that occur closer than $1/(K\bar{\nu})$.
Thus, these pairs are separated in time by on average $2/(K\bar{\nu})$
in the unperturbed trajectory. However, since they collide when they
come within $1/(K\bar{\nu})$ from one another, the susceptible pairs
in a collision event for the forward-connected and symmetric motifs
are effectively separated by the same perturbation distance as those
pairs satisfying the backward-connected motif.

\subsection{Decorrelation Cascade: derivations of $p_{\mathrm{cascade}}$ and $\lambda_{p}$}

Under the assumption of Poisson spiking activity, the probability of an irreversible
cascade, $p_{\mathrm{cascade}}$, of susceptible spike collision events
can be shown to approach unity with $K$ as follows. Equation 10 of the main
text, $\Delta t_{\mathrm{sus}}=\left(K\bar{\nu}\right)^{-1}$, arises
from the rate of susceptible spikes, $K\bar{\nu}$, being a factor
$p$ lower than that of all spikes, whose successive intervals, $\Delta t_{s}:=t_{s}-t_{s-1}$,
have average size $\left(N\bar{\nu}\right)^{-1}$. The probability
of a spike emitted by a given neuron in a time interval of size $(K\bar{\nu})^{-1}$
beginning from reference time $t=0$ is 
\begin{align*}
P(\mbox{spike in }\left[0,(K\bar{\nu})^{-1}\right]|\mathrm{{neuron\;given}}) & =\int_{0}^{(K\bar{\nu})^{-1}}\bar{\nu}e^{-\bar{\nu}t}\mbox{d}t\\
 & =1-e^{-1/K}\;.
\end{align*}
For the spiking neuron, $n_{s}$, whose spike time has jumped by $\Delta t_{\mathrm{jump}}$
due to the critical perturbation, we apply this result to the set
of its post-synaptic neurons whose activity is assumed mutually independent,
and which number $K$ on average. Defining $p_{\mathrm{post}}$ as
the probability that any of these postsynaptic neurons spike in the
window $\left[t_{s},t_{s}+(K\bar{\nu})^{-1}\right]$ across which
the spike of $n_{s}$ has jumped, 
\begin{align*}
p_{\mathrm{post}} & =\left[\sum_{\left\{ m|A_{mn_{s}}=1\right\} }P(\mbox{spike in }\left[t_{s},t_{s}+(K\bar{\nu})^{-1}\right]|\mathrm{{from\;neuron}}\;m)\right]_{n_{s}}\\
 & =KP(\mbox{spike in }\left[0,(K\bar{\nu})^{-1}\right]|\mathrm{{neuron\;given}})\\
 & =K\left(1-e^{-1/K}\right)\\
 & =K\left(1-\left(1-\frac{1}{K}+\frac{1}{2K^{2}}+\mathcal{O}\left(K^{-3}\right)\right)\right)\\
p_{\mathrm{post}} & =1-\frac{1}{2K}+\mathcal{O}\left(K^{-2}\right)\;,
\end{align*}
and so for large $K$, another collision event becomes increasingly
certain. Since the subsequent activity of a neuron involved in a crossing
event is irreversibly altered, there are on average $\log N/\log K<N/K=1/p$
number of these events until the activities of all neurons have been
altered and so $1>p_{\mathrm{cascade}}>\left(p_{\mathrm{post}}\right)^{1/p}\approx\left(1-\frac{1}{2pN}\right)^{1/p}\to1$
in the dense thermodynamic limit ($N\to\infty$ and $p$ fixed), so
that $p_{\mathrm{cascade}}\to1$.

From this microscopic explanation of the cascade, we can derive the
rate of divergence captured by the pseudoLyapunov exponent, $\lambda_{p}$,
as follows. All the future spikes of any neuron involved in such a
crossing event are shifted by about $\Delta t_{\mathrm{jump}}$ or
more, and since this neuron discharges spikes at a rate $\bar{\nu}$
and has $K$ synaptic partners, it contributes a rate $K\bar{\nu}$
of crossing events after its first, suggesting exponential growth
in the number of neurons involved in the cascade. Ordering the sequence
in which they enter the cascade by $m$, which occur at specific times,
$t_{m}$, relative to the cascade onset, the total rate of crossing
events around $t_{m}$ is roughly $mK\bar{\nu}$. An estimate for
the interval, $t_{m}-t_{m-1}$, between successive neurons joining
the cascade is then the inverse of the rate at that event, $\left(mK\bar{\nu}\right)^{-1}$.
Using the approximation $m^{-1}\sim\log(1-\frac{1}{m})^{-1}$ valid
for $m\gg1$, we can then write $t_{m}-t_{m-1}\sim\left(K\bar{\nu}\right)^{-1}\log(1-\frac{1}{m})^{-1}$,
which can be rearranged as $m/(m-1)=e^{K\bar{\nu}(t_{m}-t_{m-1})}$.
Since each entry of a neuron into the cascade brings a constant jump
in the distance, we infer up to a scaling factor that the distance
at time $t_{m}$ is $m\propto e^{K\bar{\nu}t_{m}}$, so that the increase
in the distance is exponential with a rate of $K\bar{\nu}$, explaining
the numerical result $\lambda_{p}=\frac{1}{t_{m}}\log m\approx K\bar{\nu}$
 (see Ref. 15).

\subsection{Notes on the Derivation of $S\left(\epsilon\right)$}

The statistics of the critical perturbation strength, $\epsilon^{*}$,
determine the geometry of the tube boundary, respectively. In this
section we provide details of our calculations referred to in the
main text in pursuit of $\overbar{\epsilon^{*}}$.

\subsubsection{Deviation-rate coefficients, $a_{s}$}

\begin{figure}
\centering
\includegraphics{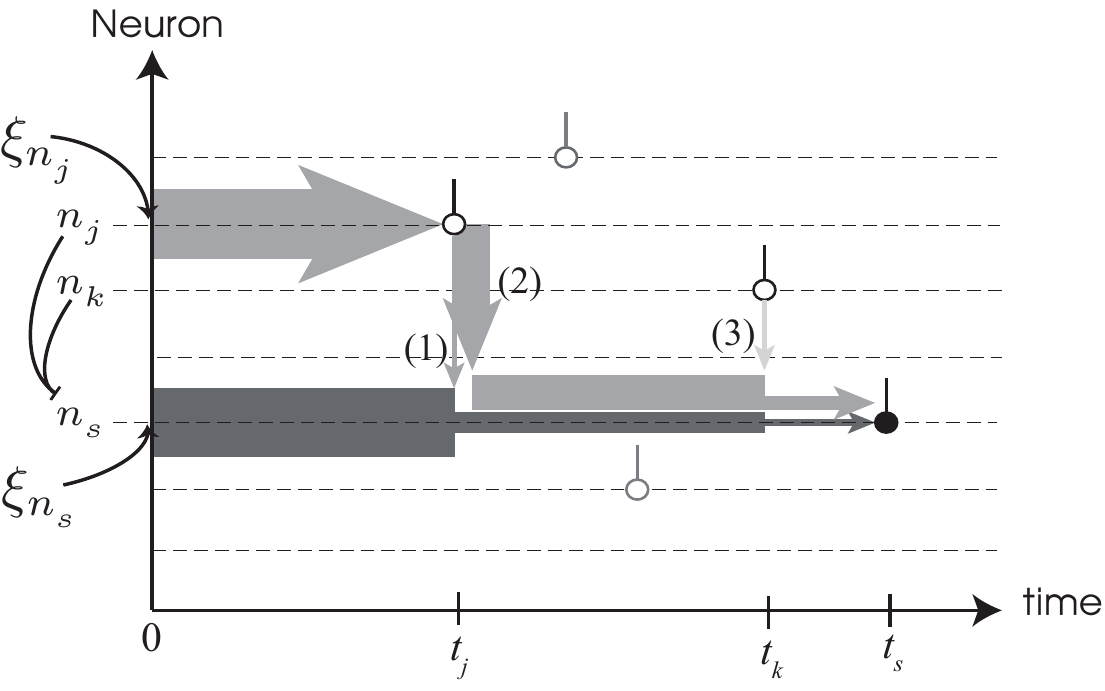}
\caption{Schematic illustration of the contributions to $a_{s}$ (equation \eqref{eq:a_Sbig-1}): 1) Direct perturbations
to neuron $n_{s}$ (dark gray) are contracted at each input spike.
2) With each input spike, the deviation from the presynaptic neuron
(light gray), scaled by $Z'(\phi)$, is added to the existing deviation
in $n_{s}$. 3) Deviations arising from (2) are also contracted with
subsequent input spikes. \label{fig:Schematic-illustration-of}}
\end{figure}

The spike time deviation, $\delta t_{s}\left(\epsilon\right):=t_{s}\left(\epsilon\right)-t_{s}\left(0\right)$,
is composed of a contribution by the direct perturbation to $n_{s}$,
and a contribution from the indirect effects of the perturbation via
deviations of the input spike times to $n_{s}$. The deviations from
both of these contributions will be contracted across subsequent input
spikes to that neuron. The derivative with respect to perturbation
strength thus consists of a differential change due to changing initial
state with fixed input spike times and a differential change due to
changing input spike times with the initial state fixed, respectively:

\begin{eqnarray*}
\frac{\mathrm{d}t_{s}}{\mathrm{d}\epsilon} & \approx & \frac{\partial t_{s}}{\partial\epsilon}+\sum_{j=1}^{s-1}\frac{\mathrm{d}t_{j}}{\mathrm{d}\epsilon}\frac{\partial t_{s}}{\partial t_{j}}\\
 & = & \frac{\mathrm{d}\phi_{n_{s}}\left(t_{1}^{-}\right)}{\mathrm{d}\epsilon}\left(\prod_{j=1}^{s-1}\frac{\partial\phi_{n_{s}}\left(t_{j}^{+}\right)}{\partial\phi_{n_{s}}\left(t_{j}^{-}\right)}\right)\frac{\partial t_{s}}{\partial\phi_{n_{s}}\left(t_{s-1}^{+}\right)}+\sum_{j=1}^{s-1}\frac{\mathrm{d}t_{j}}{\mathrm{d}\epsilon}\frac{\partial\phi_{n_{s}}\left(t_{j}^{-}\right)}{\partial t_{j}}\left(\prod_{k=j+1}^{s-1}\frac{\partial\phi_{n_{s}}\left(t_{k}^{+}\right)}{\partial\phi_{n_{s}}\left(t_{k}^{-}\right)}\right)\frac{\partial t_{s}}{\partial\phi_{n_{s}}\left(t_{s-1}^{+}\right)}\\
 & = & \left(\frac{\xi_{n_{1}}}{\sqrt{N}}\right)\left(\prod_{j=1}^{s-1}\left(1+d_{\phi_{s}^{j}}\right)^{A_{n_{s}n_{j}}}\right)\left(-T_{free}\right)+\sum_{j=1}^{s-1}\left(-\frac{1}{T_{free}}A_{n_{s}n_{j}}d_{\phi_{s}^{j}}\right)\left(\prod_{k=j+1}^{s-1}\left(1+d_{\phi_{s}^{k}}\right)^{A_{n_{s}n_{k}}}\right)\left(-T_{free}\right)\frac{\mathrm{d}t_{j}}{\mathrm{d}\epsilon}\\
\frac{\mathrm{d}t_{s}}{\mathrm{d}\epsilon} & = & \frac{-T_{free}}{\sqrt{N}}\xi_{n_{1}}\left(\prod_{j=1}^{s-1}\left(1+d_{\phi_{s}^{j}}\right)^{A_{n_{s}n_{j}}}\right)+\sum_{j=1}^{s-1}A_{n_{s}n_{j}}d_{\phi_{s}^{j}}\left(\prod_{k=j+1}^{s-1}\left(1+d_{\phi_{s}^{k}}\right)^{A_{n_{s}n_{k}}}\right)\frac{\mathrm{d}t_{j}}{\mathrm{d}\epsilon}
\end{eqnarray*}
where $d_{\phi_{s}^{j}}$ is shorthand for the derivative of the PRC,
\begin{align*}
d_{\phi_{s}^{j}} & :=Z'(\phi_{n_{s}}(t_{j}))\;,
\end{align*}
evaluated at the phase of the $n_{s}$ neuron at the time of the $j^{\mbox{th}}$
spike in the network spike sequence, and where the perturbation direction
vector, $\vec{\xi}$, is not normalized but explicitly divided by
$\sqrt{N}$, preserving the $\mathcal{O}\left(1/\sqrt{N}\right)$-scaling
of a unit vector. Dividing through by $\frac{-T_{free}}{\sqrt{N}}$,
and rescaling the perturbation to $\tilde{\epsilon}=\frac{-T_{free}}{\sqrt{N}}\epsilon$,
we obtain 
\begin{equation}
\delta t_{s}\left(\epsilon\right)=\frac{\mathrm{d}t_{s}}{\mathrm{d}\tilde{\epsilon}}\frac{\mathrm{d}\tilde{\epsilon}}{\mathrm{d}\epsilon}\epsilon=-\frac{T_{free}}{\sqrt{N}}a_{s}\epsilon\;,
\end{equation}
as quoted in the main text (equation 10), where the recursively defined
sequence of dimensionless susceptibilities $a_{s}$ are given by
\begin{eqnarray}
a_{1} & =\frac{\mathrm{d}t_{1}}{\mathrm{d}\tilde{\epsilon}}= & \xi_{n_{1}}\nonumber \\
a_{2} & =\frac{\mathrm{d}t_{2}}{\mathrm{d}\tilde{\epsilon}}= & \xi_{n_{2}}\left(1+d_{\phi_{2}^{1}}\right)^{A_{n_{2}n_{1}}}+A_{n_{2}n_{1}}d_{\phi_{2}^{1}}a_{1}\nonumber \\
a_{3} & =\frac{\mathrm{d}t_{3}}{\mathrm{d}\tilde{\epsilon}}= & \xi_{n_{3}}\left(1+d_{\phi_{3}^{1}}\right)^{A_{n_{3}n_{1}}}\left(1+d_{\phi_{3}^{2}}\right)^{A_{n_{3}n_{2}}}+A_{n_{3}n_{2}}d_{\phi_{3}^{2}}a_{2}+A_{n_{3}n_{1}}d_{\phi_{3}^{1}}\left(1+d_{\phi_{3}^{2}}\right)^{A_{n_{3}n_{2}}}a_{1}\nonumber \\
\vdots & \vdots & \vdots\nonumber \\
a_{s}: & =\frac{\mathrm{d}t_{s}}{\mathrm{d}\tilde{\epsilon}}= & \underbrace{{\xi_{n_{s}}\prod_{j=1}^{s-1}\left(1+d_{\phi_{s}^{j}}\right)^{A_{n_{s}n_{j}}}}}_{(1)}+\sum_{j=1}^{s-1}\underbrace{{A_{n_{s}n_{j}}d_{\phi_{s}^{j}}a_{j}}}_{(2)}\underbrace{{\left(\prod_{k=j+1}^{s-1}\left(1+d_{\phi_{s}^{k}}\right)^{A_{n_{s}n_{k}}}\right)}}_{(3)}\;.\label{eq:a_Sbig-1}
\end{eqnarray}
The three numbered contributions in equation \ref{eq:a_Sbig-1} are shown
and described in a schematic illustration in Fig. \ref{fig:Schematic-illustration-of}.
equation \ref{eq:a_Sbig-1} reflects how arbitrary connectivity and
single neuron dynamics (via $Z(\phi)$) enter into the perturbed
spiking activity. We validated this expression (see Fig. 4 in main
text) and confirmed this exact linear dependence of $\delta t_{s}\left(\epsilon\right)$
on $\epsilon$ with direct numerical simulations over multiple values
of the perturbation strength, $\epsilon$ and out to $s=4000$. We
note that when a pair of non-susceptible spikes collide $a_{s}$ and $a_{s-1}$ exchange values, so that we occasionally
observed a piece-wise dependence of $a_{s}$ on $\epsilon$, though
this occurred infrequently. There were also a small number of cases
where we observed a small quadratic component to $\delta t_{s}\left(\epsilon\right)$,
noticeable over the range of perturbation strength within the local tube.

\subsubsection{Derivation Notes}

In the main text, we state the density for the perturbed intervals
in terms of the ingredients appearing in their linear approximation,
equation (16), 

\begin{equation}
\rho_{T}=\rho\left(\left\{ \Delta a_{s}\right\} ,\left\{ \Delta t_{s}\right\} ,M,\vec{\phi}_{0}\bigg\vert\left(A_{mn}\right),\vec{\xi}\right)\rho\left(\vec{\xi}\right)P_{A}\left(\left(A_{mn}\right)\right)
\end{equation}
where the susceptibilities, $\left\{ a_{s}\right\} $, describing
the rate of spike time deviation as a function of perturbation strength
are given by equation \ref{eq:a_Sbig-1} and so depend on elements of the
connectivity, $A=\left(A_{mn}\right)$, the perturbation direction, 
$\vec{\xi}$, and $d_{\phi_{s}^{j}}$, the phase response curve evaluated
at the phase of the $n_{s}$ neuron at the time of the $j^{\mbox{th}}$
spike in the network spike sequence. The state being
perturbed at $t=0$, $\vec{\phi}_{0}$, is an equilibriated state
whose probability density function depends in general on the realization of the connectivity,
$A=\left(A_{mn}\right)$. For large, sparse connectivities, however,
the self-averaging properties of $A$ leave the invariant density
of states $\rho\left(\vec{\phi}_{0}\right)$ dependent only on the parameters
of the connectivity ensemble and not the particular realization. A
closed form for this density has been previously derived (see Ref. 17), though
we will not need it here, since the distributions of unperturbed intervals
arising from $\vec{\phi}_{0}$ is explicitly considered and the dependence
of $d_{\phi_{s}^{j}}$ on $\phi_{s}^{j}$ becomes negligible at large $K$ (equation 7).

\begin{figure}
\centering
\includegraphics{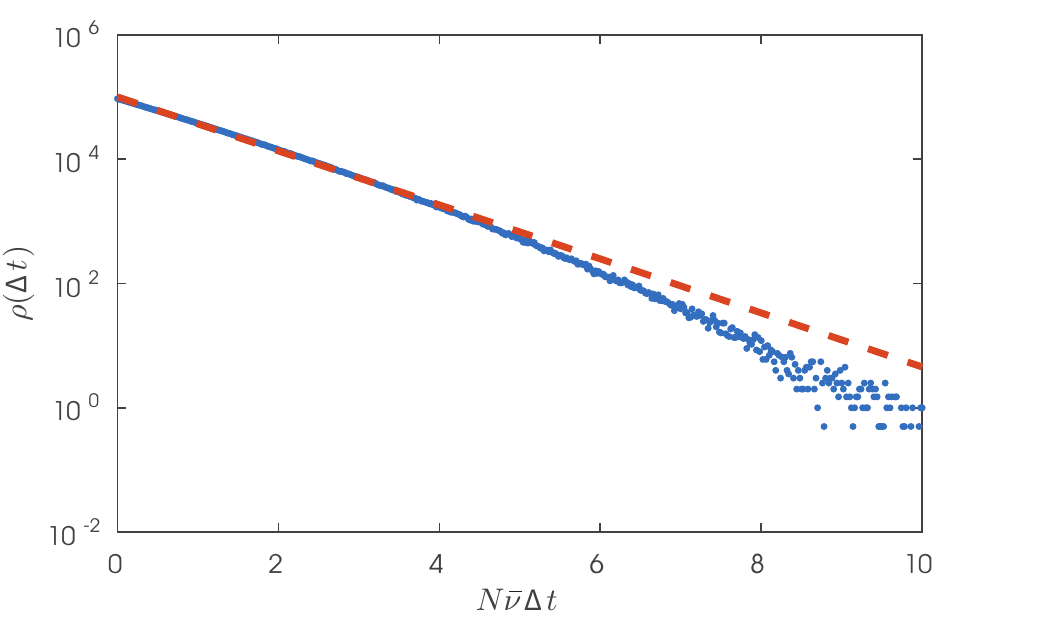}
\caption{\label{fig:Spiketime-intervals} Network spike time interval probability
density,, $\rho(\Delta t)$. It is distributed exponentially ($N=10^{4}$, $\bar{\nu}=10\;\mathrm{{Hz}}$,
$10^{7}$ network intervals). Dashed line is the prediction, $\rho\left(\Delta t\right)=N\bar{\nu}e^{N\bar{\nu}\Delta t}$.
Note that the abscissa is scaled by $N\bar{\nu}$.}
\end{figure}

In a diffusion approximation, applicable to large, sparse graphs,
the inputs to a unit are negligibly correlated. In particular, the sum of many, weakly correlated inputs,
is Poisson (see Ref. 21). Each of the set of unperturbed inter-spike
intervals, $\left\{ \Delta t_{s}\right\} $, of the compound spike
sequence obeys a distribution that rapidly approaches with increasing $N$ the same exponential
form with rate $N\bar{\nu}$, $\rho_{t}(\Delta t_{s})=N\bar{\nu}e^{-N\bar{\nu}\Delta t_{s}}$
for all $s$ (see Ref. 21). The distribution of $\left\{ \Delta t_{s}\right\} $
is then 
\begin{equation}
\rho(\left\{ \Delta t_{s}\right\} )=\prod_{s=2}^{M}\rho_{t}(\Delta t_{s})\;.
\end{equation}
In Fig.\ref{fig:Spiketime-intervals}, we show $\rho(\left\{ \Delta t_{s}\right\} )$.
 It is indeed exponential up to finite-sampling effects. This validates
the assumption of exponential interval statistics.

\paragraph{Simplifying $\Delta a_{s}$}

$\Delta a_{s}$ simplifies in three ways. The size of
indirect effects (the second term in equation \ref{eq:a_Sbig-1}) can be
ignored for large $K$, since they additionally contain $d_{\phi_{s}^{j}}\propto K^{-1}$
as a factor. Thus,
\begin{equation}
a_{s}\approx\xi_{n_{s}}\prod_{j=1}^{s-1}\left(1+d_{\phi_{s}^{j}}\right)^{A_{n_{s}n_{j}}}\;.
\end{equation}
The small synaptic strength linearizes $Z\left(\phi\right)$ for $K\gg1$
so that $d_{\phi_{s}^{j}}\approx-d$ with $d:=(K\bar{\nu}\tau)^{-1}>0$
(equation 7) so $a_{s}$ no longer depends on the distribution of states.
Third, for non-small $s$ a fraction $p$ of the earlier spikes $\{1,\dots,s-1\}$
are from neurons presynaptic to $n_{s}$ so that $a_{s}\approx\xi_{n_{s}}\left(1-d\right)^{\sum_{j=1}^{s-1}A_{n_{s}n_{j}}}\approx\xi_{n_{s}}\left(1-d\right)^{ps}=\xi_{n_{s}}\left(\left(1-(\bar{\nu}\tau)^{-1}/K\right)^{K}\right)^{\frac{s}{N}}\approx\xi_{n_{s}}e^{-pds}$
for $K\gg1$. Thus, $\Delta a_{s}\approx e^{-pd(s+1)}\xi_{n_{s+1}}-e^{-pds}\xi_{n_{s}}=e^{-pds}\left(\xi_{n_{s+1}}-\xi_{n_{s}}\right)$,
since $e^{-pd}\approx1$ for $N\gg1$. We note that $-pds\approx\lambda t$
where $\lambda=-\tau^{-1}$ serves here as an estimate for mean Lyapunov
exponent, $\lambda_{\mathrm{mean}}$, at large $K$, formally calculated
in Ref. 15.

$\sigma_{\xi}$ then determines the numeric prefactor in the standard
deviation of $\Delta a_{s}$, $\sigma_{\Delta a_{s}},$ and so can
be set to make this prefactor unity. $\rho(\xi)$ was chosen as a
centered normal distribution in order to generate isotropic perturbation
directions. The difference of two independent centered normal random
variables has 0 mean and twice the variance. Thus, $\sigma_{\Delta a_{s}}=\sqrt{2}\sigma_{\xi}e^{-pds}$. 
We also note that $\left[a_{s}\right]_{\rho(\vec{\xi})}=0$ under
the negligible serial correlation assumption used in the main text,
since $\left[\xi_{n_{s}}\right]_{\rho(\vec{\xi})}=0$ for the isotropic
perturbation direction distributions used here. 
It the selection of only positive $\Delta a_{s}$ that gives a finite expectation.

\paragraph{Evaluating the expectation in $S_{s}\left(\epsilon\right)$}

We note that for a centered, normally-distributed variable, $x$,
the corresponding distribution over only the positive range is
\begin{equation}
\mathcal{N}^{+}\left(0,\sigma_{x}^{2}\right)=2\Theta\left(x\right)\mathcal{N}\left(0,\sigma_{x}^{2}\right)\;.
\end{equation}
Applying this result to $\rho_a(\Delta a_{s})$, equation 19 is then evaluated as

\begin{eqnarray*}
S_{s}\left(\epsilon\right) & = & \left[\Theta\left(\Delta t-\frac{T_{free}}{\sqrt{N}}\Delta a_{s}\epsilon\right)^{A_{mn}}\right]_{\rho_{t}\left(\Delta t\right)\rho_{a}\left(\Delta a_{s}\right)P_{A_{mn}}\left(A_{mn}\right)}\\
 & = & (1-p)+p\int_{0}^{\infty}2\rho_{a}\left(\Delta a_{s}\right)\mathrm{d}\Delta a_{s}\int_{0}^{\infty}\rho_{t}\left(\Delta t\right)\mathrm{d}\Delta t\left(\Theta\left(\Delta t-\frac{T_{free}}{\sqrt{N}}\Delta a_{s}\epsilon\right)\right)\\
 & = & (1-p)+p\int_{0}^{\infty}2\frac{1}{\sqrt{2\pi}\sigma_{\Delta a_{s}}}e^{-\frac{\left(\Delta a_{s}\right)^{2}}{2\sigma_{\Delta a_{s}}^{2}}}\mathrm{d}\Delta a_{s}\left(\int_{\frac{T_{free}}{\sqrt{N}}\Delta a_{s}\epsilon}^{\infty}\mathrm{d}\Delta t\frac{1}{\overbar{\Delta t}}e^{-\frac{\Delta t}{\overbar{\Delta t}}}\right)\\
 & = & (1-p)+p\int_{0}^{\infty}2\frac{1}{\sqrt{2\pi}\sigma_{\Delta a_{s}}}e^{-\frac{\left(\Delta a_{s}\right)^{2}}{2\sigma_{\Delta a_{s}}^{2}}-\frac{T_{free}}{\overbar{\Delta t}}\frac{\epsilon}{\sqrt{N}}\Delta a_{s}}\mathrm{d}\Delta a_{s}\\
 & = & (1-p)+p\frac{2}{\sqrt{\pi}}e^{x_{s}^{2}}\int_{0}^{\infty}e^{-(z_{s}+x_{s})^{2}}\mathrm{d}z_{s}\\
 & = & (1-p)+p\frac{2}{\sqrt{\pi}}e^{x_{s}^{2}}\int_{x_{s}}^{\infty}e^{-y_{s}^{2}}\mathrm{d}y_{s}\\
 & = & (1-p)+p\frac{2}{\sqrt{\pi}}e^{x_{s}^{2}}\left(\frac{\sqrt{\pi}}{2}-\int_{0}^{x_{s}}e^{-y_{s}^{2}}\mathrm{d}y_{s}\right)\\
S_{s}\left(\epsilon\right) & = & (1-p)+p\mathrm{Erfcx}\left(x_{s}\right)
\end{eqnarray*}
with $z_{s}=\Delta a_{s}/\left(\sqrt{2}\sigma_{\Delta a_{s}}\right)$,
$y_{s}=z_{s}+x_{s}$, and $x_{s}=\frac{T_{free}}{\overbar{\Delta t}}\frac{\epsilon}{\sqrt{N}}\sigma_{\Delta a_{s}}$.

\subsubsection{Derivation of scaling for non-negligible activity correlations}

In the main text, we give a derivation of $\overbar{\epsilon^{*}}$
relying on the assumption that serial correlations are negligible
and so each interval can be taken, conditioned on $s$, as an independent
sample. This seemingly strong assumption is supported for large networks
in the asynchronous, irregular activity state by the fact that there is a negligible probability
that nearby spikes arise from neurons separated by few connections.
In particular, the serial interval correlations $C\left(s\right)=\left[\Delta t_{s'}\Delta t_{s'+s}\right]_{s'}\to(N\bar{\nu})^{-2}\delta(s)$
in the thermodynamic limit, $N\to\infty$. In this section, we present
a general approach to obtaining the scaling of the average critical
diameter, $\overbar{\epsilon^{*}}$, free of the assumption of no
serial correlation of $s$-indexed quantities.

$\mathbb{1}_{\mathrm{FT}}\left(\epsilon\right)$ can be exactly re-expressed
using the set of smallest positive zeros, $\epsilon_{s}^{*}=\frac{\sqrt{N}}{T_{free}}\frac{\Delta t_{s}}{\Delta a_{s}}>0$,
of the linearized $\Delta t_{s}\left(\epsilon\right)$ where $A_{n_{s}n_{s+1}}=1$.
By writing the connection motif condition as  $b_{s}=1$, for binary-valued function
$b_{s}$, we can incorporate other divergence event flags (e.g. $b_{s}=A_{n_{s+1}n_{s}}$
flags the forward connected motif). We then write the indicator function
as $\mathbb{1}_{\mathrm{FT}}\left(\epsilon\right)\equiv\lim_{T\to\infty}\mathbb{1}_{T}(\epsilon)$,
with 
\begin{equation}
\mathbb{1}_{T}(\epsilon):=\prod_{s=2}^{M}\Theta\left(\epsilon_{s}^{*}-\epsilon\right)^{b_{s}}\;,\label{eq:prodTheta-1}
\end{equation}
the convention $0^{0}=1$, and $M$ the number of spikes observed
in the time window $\left[0,T\right]$. The procedure to obtain a
scale from $S\left(\epsilon\right)=\left[\mathbb{1}_{\mathrm{FT}}\left(\epsilon\right)\right]_{\rho(A,\vec{\phi_{0}},\vec{\xi})}$
is as follows.

We introduce a scale into $\mathbb{1}_{T}(\epsilon)$ by (1) multiplying
each $s$-indexed factor of equation \ref{eq:prodTheta-1} with $k_{s}\left(\epsilon_{s}^{*}-\epsilon\right)$,
where the constants $k_{s}$ are chosen depending on the form of $\rho_T$
such that the expectation remains unchanged; and (2) excluding with
a new power, $b_{s}^{+}=b_{s}\Theta\left(\frac{\mathrm{d}(\delta t_{s-1}\left(\epsilon\right)-\delta t_{s}\left(\epsilon\right))}{\mathrm{d}\epsilon}\right)$,
values of $s$ corresponding to growing intervals, since these cannot
generate a divergence event. We denote this scale-dependent indicator
function, 
\begin{align}
i_{T}\left(\epsilon\right) & :=\prod_{s=2}^{M}\left(k_{s}\left[\epsilon_{s}^{*}-\epsilon\right]_{+}\right)^{b_{s}^{+}}\,,\label{eq:scaledver}
\end{align}
where $\left[x\right]_{+}\equiv x\Theta(x)$. This transformation
gives $i_{T}\left(\epsilon\right)$ a scale and thus has the effect
of lowering the leading order of the expectation of around $\epsilon=0$
from $\mathcal{O}\left(\epsilon^{2}\right)$ to $\mathcal{O}\left(\epsilon\right)$. 

$S\left(\epsilon\right)$ integrates over a constant density of decorrelation
events, making it linear at small
$\epsilon$. The expectation of the Taylor-expanded function there
is 
\begin{equation}
\left[i_{T}\left(\epsilon\right)\right]_{\rho_{T}}=\left[\prod_{s=2}^{M}\left(k_{s}\epsilon_{s}^{*}\right)^{b_{s}^{+}}\right]_{\rho_{T}}-\left[\sum_{s=2}^{M}\left(\prod_{s'\neq s}^{M}\left(k_{s'}\epsilon_{s'}^{*}\right)^{b_{s'}^{+}}\right)b_{s}^{+}k_{s}\right]_{\rho_{T}}\epsilon+\mathcal{O}\left(\epsilon^{2}\right)\;,
\end{equation}
so that the characteristic scale is, 
\begin{equation}
\overbar{\epsilon^{*}}:=\lim_{T\to\infty}\frac{\frac{\sqrt{N}}{T_{free}}\left[\prod_{s=2}^{\left\lfloor T/\overbar{\Delta t}\right\rfloor }\left(k_{s}\frac{\Delta t_{s}}{\Delta a_{s}}\right)^{b_{s}^{+}}\right]_{\rho_{T}^{+}}}{\sum_{s=2}^{\left\lfloor T/\overbar{\Delta t}\right\rfloor }\left[\left(\prod_{s'\neq s}^{\left\lfloor T/\overbar{\Delta t}\right\rfloor }\left(k_{s'}\frac{\Delta t_{s'}}{\Delta a_{s'}}\right)^{b_{s'}^{+}}\right)b_{s}^{+}k_{s}\right]_{\rho_{T}^{+}}}\label{eq:mostgenEps}
\end{equation}
where $\left\lfloor T/\overbar{\Delta t}\right\rfloor \approx M$,
with average interval size $\overbar{\Delta t}=(N\bar{\nu})^{-1}$ ($\left\lfloor x\right\rfloor$ denotes the largest integer less than $x$).
In this derivation, no assumptions about the network graph or the
unit dynamics have been made, and we expect equation \ref{eq:mostgenEps}
to hold generally for phase spaces partitioned by divergence events.
The $k_{s}$ depend on the serial interval correlations and on which
of the invariances of the expectation operation, arising from the
form of $\rho_T$, is exploited
when introducing a scale to $\mathbb{1}_{T}(\epsilon)$. 

For the networks considered here, the expectation of negative powers
of $\Delta t_{s}$ and $\Delta a_{s}$ diverge. With $k_{s}=\Delta a_{s}/\overbar{\Delta t}$,
the expectations in equation\ref{eq:mostgenEps} are of positive powers
only and so remain finite. This choice of $k_{s}$ also makes the
resulting expectation of equation \ref{eq:prodTheta-1} equivalent to that
of equation \ref{eq:scaledver}. To see this, consider first only the exponentially-distributed
$\Delta t_{s}$, with $\Delta a_{s}$ fixed. In this case, the invariance
can be simply obtained by integrating by parts. For example, $\int_{c}^{\infty}\frac{(\Delta t-c)}{\overbar{\Delta t}}\frac{1}{\overbar{\Delta t}}\mbox{Exp}\left[-\Delta t/\overbar{\Delta t}\right]\mbox{d}\Delta t=\int_{c}^{\infty}\frac{1}{\overbar{\Delta t}}\mbox{Exp}\left[-\Delta t/\overbar{\Delta t}\right]\mbox{d}\Delta t$
for any constant $c>0$. Indeed, $\rho_T$
admits the desired invariance of the expectation captured in the relation
$\left[\Theta\left(\epsilon_{s}^{*}-\epsilon\right)\frac{\Delta t_{s}\left(\epsilon\right)}{\overbar{\Delta t}}\right]_{\rho_{t}\left(\Delta t\right)\rho_{a}\left(\Delta a_{s}\right)}=\left[\Theta\left(\epsilon_{s}^{*}-\epsilon\right)\right]_{\rho_{t}\left(\Delta t\right)\rho_{a}\left(\Delta a_{s}\right)}$
so that with this choice of $k_{s}$ the value of $\left[i_{T}\left(\epsilon\right)\right]_{\rho_{T}}$
remains the same as $\left[\mathbb{1}_{T}\left(\epsilon\right)\right]_{\rho_{T}}$. 

We can evaluate equation \ref{eq:mostgenEps} for the networks considered
here. With $k_{s}=\Delta a_{s}/\overbar{\Delta t}$, equation \eqref{eq:scaledver} becomes
\begin{equation}
i_{T}\left(\epsilon\right)=\prod_{s=2}^{M}\left(\Theta\left(\epsilon_{s}^{*}-\epsilon\right)\frac{\Delta t_{s}\left(\epsilon\right)}{\overbar{\Delta t}}\right)^{b_{s}^{+}}
\end{equation}
and $\left[i_{T}\left(0\right)\right]_{\rho_{T}}=1$ as required. 
The negligible serial correlations permit commutation of the
expectation and product so that equation \ref{eq:mostgenEps} simplifies to
\begin{align*}
\overbar{\epsilon^{*}} & \approx\frac{1}{C}\left(\sum_{s=2}^{\infty}\left[b_{s}^{+}\Delta a_{s}\right]_{\rho_{t}\left(\Delta t\right)\rho_{a}\left(\Delta a_{s}\right)P_{A_{mn}}\left(A_{mn}\right)}\right)^{-1}\;,
\end{align*}
which can be evaluated using the steps in the derivation presented
in the main text, and gives the same result (equation 22).

\subsection{Non-folded phase-space representation of the flux tube partition}

\begin{figure}
\centering
\includegraphics{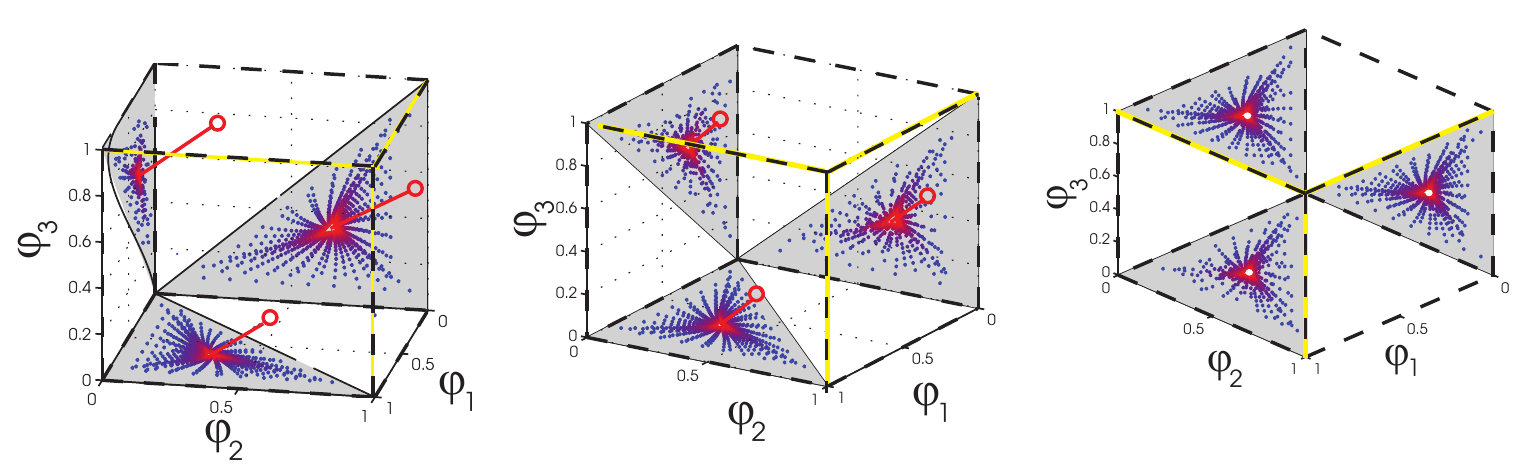}
\caption{\label{fig:A-fully-connected-3-neuron}A fully-connected 3-neuron
network phase space viewed from rotated perspectives (from left to
right) so that the main diagonal aligns perpendicular to the page.
Iterated on the reset manifold, all states are attracted (blue to
red) in time to a unique trajectory (red line), emitting spikes
on the threshold manifold (the red outlined dots). The susceptible edges (yellow)
and their back projections (black-dashed) form the
basin boundaries of the flux tube partition. The two flux tube attractors
in this 3D phase-space correspond to two unique periodic spike index sequences
of the dynamics, $\dots n_{1}n_{2}n_{3}\dots$ and $\dots n_{2}n_{1}n_{3}\dots$.}
\end{figure}

For concreteness, in Fig. \ref{fig:A-fully-connected-3-neuron} we
show an actual non-folded phase-space representation of the flux tube
partition, here for a network of three neurons, each connected to
the two others.

\let\bibsection\relax

\newpage

